%% file: main.tex
\documentclass[sigconf,balance=false]{acmart}
\usepackage{popets}

\usepackage{algorithm}
\usepackage{algpseudocode}
\usepackage{amsmath}
\usepackage{tikz}
\usetikzlibrary{calc}
\usepackage{pifont}
\usepackage{array}
\usepackage{adjustbox}
\usepackage{stfloats}
\usepackage{graphicx}
\usepackage{subcaption}
\usepackage{multirow}
\usepackage{xcolor}
\usepackage{enumitem}

\setcopyright{popets}
\copyrightyear{YYYY}

\acmYear{YYYY}
\acmVolume{YYYY}
\acmNumber{X}
\acmDOI{XXXXXXX.XXXXXXX}
\acmISBN{}
\acmConference{Proceedings on Privacy Enhancing Technologies}
\begin{document}

%%%%%%%%%%%%%%%% Custom Commands %%%%%%%%%%%%%%%%%%%%%%%%%%
\newcommand{\ibrahim}[1]{{\footnotesize\color{orange}[Ibrahim: #1]}}
\newcommand{\bo}[1]{{\footnotesize\color{magenta}[Bo: #1]}}
\newcommand{\matt}[1]{{\footnotesize\color{blue}[Matt: #1]}}
\newcommand{\bdj}[1]{{\footnotesize\color{red}[Brendan: #1]}}

\newcommand{\xmark}{\ding{55} }
\newcommand{\emptycircle}{\ding{109} }
\newcommand{\filledcircle}{\ding{108} }

\newcommand{\frameworkname}{\textsc{EvaluatAR}}

% Place a mark on the baseline of the current line (no new line)
\newcommand{\AlgMark}[1]{%
  \tikz[overlay,remember picture,baseline] \coordinate (#1);%
}
% Draw a full-width dotted rectangle between two marks, with explicit padding
% \DrawTightBlock{color}{start}{end}{leftpad}{rightpad}{topPad}{botPad}{yshift}
\newcommand{\DrawTightBlock}[8]{%
\begin{tikzpicture}[overlay,remember picture]
% left/right x positions (full line)
\coordinate (L) at ($ (#2) + (-#4,0) $);
\coordinate (R) at ($ (#2) + (\linewidth + #5,0) $);
% top/bottom y positions from start/end marks, with explicit pads
\coordinate (T) at ($ (#2) + (0,#6) $);
\coordinate (B) at ($ (#3) + (0,-#7) $);
% rectangle corners
\coordinate (TL) at ($ (L |- T) + (0,#8) $);
\coordinate (BR) at ($ (R |- B) + (0,#8) $);
\draw[
  color=#1,
  densely dotted,
  line width=0.4pt,
  rounded corners=1pt
] (TL) rectangle (BR);
\end{tikzpicture}%
}

%%%%%%%%%%%%%%%% Title %%%%%%%%%%%%%%%%%%%%%%%%%%
\title[\frameworkname]{\frameworkname: A Cross-Device Evaluation Framework for Rapid Prototyping of Bystander PETs in AR}

%%%%%%%%%%%%%%%% Authors' Info %%%%%%%%%%%%%%%%%

% \author{Syed Ibrahim Mustafa Shah Bukhari}
% \orcid{0009-0004-1174-8660}
% \affiliation{%
%   \institution{Virginia Tech}
%   \city{Blacksburg}
%   \state{VA}
% \email{simsb@vt.edu}
\author{Syed Ibrahim Mustafa Shah Bukhari}
\orcid{0009-0004-1174-8660}
\affiliation{%
  \institution{Virginia Tech}
  \city{}
  \state{}
  \country{}}
\email{simsb@vt.edu}

% \author{Matthew Corbett}
% \affiliation{%
%  \institution{Army Cyber Institute at West Point}
%  \city{West Point}
%  \state{NY}
% \email{matthew.corbett@westpoint.edu}
\author{Matthew Corbett}
\affiliation{%
 \institution{Army Cyber Institute at West Point}
 \city{}
 \state{}
 \country{}}
\email{matthew.corbett@westpoint.edu}

\author{Bo Ji}
\affiliation{%
  \institution{Virginia Tech}
  \city{}
  \state{}
  \country{}}
\email{boji@vt.edu}

\author{Brendan David-John}
\affiliation{%
  \institution{Virginia Tech}
  \city{}
  \state{}
  \country{}
}
\email{bmdj@vt.edu}

\renewcommand{\shortauthors}{Bukhari et al.}

\begin{abstract}

Augmented Reality (AR) headsets continuously sense their surroundings, capturing nearby bystanders and raising privacy risks. Visual bystander privacy-enhancing technologies (PETs) mitigate this risk by detecting bystanders in egocentric scene views and applying privacy transformations (e.g., obfuscation). However, traditional PET evaluation is human-dependent, high-overhead, and device-specific, making it difficult to reproduce across devices. We present \frameworkname, a cross-device evaluation framework for rapid prototyping at the early stage of PET evaluation. Our framework enables controlled replication of experimental conditions by standardizing PET inputs (sensor data and visual stimuli) and outputs through a record-replay workflow. We validate \frameworkname\ through three case studies on HoloLens~2, Magic Leap~2, and Meta Quest~3 across implicit (continuous, context-driven) and explicit (intent-driven) PETs:  
(1) cross-device replay of inputs to a PET to reveal device-specific privacy-performance trade-offs; 
(2) generalizability of the same framework workflow across implicit and explicit PET design categories; 
and
(3) replay of privacy-relevant edge cases to diagnose failures and validate PET modifications, yielding an improvement over the state-of-the-art baseline. 
These results demonstrate \frameworkname's support for rapid, iterative PET development to advance reproducible cross-device evaluation of bystander PETs at a critical moment in the emergence of ubiquitous AR.
% }

\end{abstract}

%%%%%%%%%%%%%%%% Keywords %%%%%%%%%%%%%%%%%%%%%%%
% \keywords{datasets, neural networks, gaze detection, text tagging}

\maketitle

%%%%%%%%%%%%%%%% Introduction %%%%%%%%%%%%%%%%%%%
\input{TextFiles/New_Introduction.tex}

%%%%%%%%%%%%%%%% Related Works %%%%%%%%%%%%%%%%%%
\input{TextFiles/New_RelatedWork.tex}

%%%%%%%%%%%%%%%% Framework %%%%%%%%%%%%%%%%%%%%
\input{TextFiles/New_Framework.tex}

%%%%%%%%%%%%%%%% Framework %%%%%%%%%%%%%%%%%%%%
\input{TextFiles/New_FrameworkImplementation.tex}

%%%%%%%%%%%%%%%% Methodology %%%%%%%%%%%%%%%%%%%%
\input{TextFiles/New_CaseStudies.tex}

%%%%%%%%%%%%%%%% Discussion %%%%%%%%%%%%%%%%%%%%%
\input{TextFiles/New_Discussion.tex}

%%%%%%%%%%%%%%%% Conclusion %%%%%%%%%%%%%%%%%
\input{TextFiles/New_Conclusion.tex}

\begin{acks}

The authors acknowledge support from the Commonwealth Cyber Initiative Southwest Virginia Node, the National Science Foundation under Award No. CNS-2350116, and the Center for Human-Computer Interaction at Virginia Tech. Any opinions, findings, and conclusions or recommendations expressed in this material are those of the author(s) and do not necessarily reflect the views of the Commonwealth Cyber Initiative, the National Science Foundation, or VT's Center for Human-Computer Interaction. 

Moreover, the authors would like to acknowledge the use of ChatGPT to revise the text throughout all sections of the paper to correct typos, grammatical errors, and awkward phrasing.

\end{acks}

%%%%%%%%%%%%%%%% Bibliography %%%%%%%%%%%%%%%%%%%%%%%%%%
\bibliographystyle{ACM-Reference-Format}
\bibliography{sample-base}

%%%%%%%%%%%%%%%% Appendix %%%%%%%%%%%%%%%%%%%%%%%%%%
\appendix

% \newpage
\section{\frameworkname's Algorithm}\label{appendix-evaluataralgo}
Alg.~\ref{alg:evaluatar} visualizes the control loop of \frameworkname. In \texttt{Collect} mode, \frameworkname\ records time-synchronized sensor inputs and runtime measurements. In \texttt{Replay} mode, it (1) establishes alignment using the reference marker, (2) disables alignment components once positioned to reduce overhead, and (3) replays logged inputs by selecting the appropriate log entry as a function of elapsed replay time. This time-based replay mechanism supports reproducible alignment between replayed inputs and the stimulus across headsets with different processing rates.

\begin{algorithm}[h]
\caption{\textsc{EvaluatAR} control loop.}
\label{alg:evaluatar}
\begin{algorithmic}[1]

\State \textbf{Parameters:} PositioningCube, AlignmentVisualizer, QRDetector, $H_t$, $Q_t$, ReplayLogsPath, LogData, CameraFrame, Mode $\in \{Collect, Replay\}$
\State logBuffers, elapsedT = 0
\State isTogglePressed, isRefImgCaptured = false

\If{Mode $== Replay$}
    \State $logBuffers \gets Load(ReplayLogsPath)$
\Else
    \State Disable PositioningCube, AlignmentVisualizer
\EndIf

\While{True}
    \State Calculate FPS
    \If{isTogglePressed}
        \State elapsedT $\gets StopWatch.Start$
        \State isTogglePressed $\gets false$
    \EndIf
    
    \If{Mode $== Collect$}
        \State $logBuffers \gets (time, elapsedT, FPS, LogData)$       
    \Else
        \If{$H_t \in PositioningCube.Bounds$}
            \If{isRefImgCaptured}
                \State Disable QRDetector
            \Else
                \State Save CameraFrame
                \State isRefImgCaptured $\gets true$
            \EndIf
        \Else
            \State UpdatePositioningCubePose($Q_t, logBuffers.QRPose$)
        \EndIf

        \If{isTogglePressed}
            \For{j $\gets 1$ \textbf{to} j $\leq LogData.Count$}
                \If{elapsedT $==$ logBuffers[j].elapsedT}
                    \State Replay logBuffers[j].elapsedT
                \EndIf
                \If{elapsedT $<$ logBuffers[j].elapsedT}
                    \State Replay logBuffers[j$- 1$].elapsedT
                \EndIf
            \EndFor
        \EndIf
    \EndIf
\EndWhile
\end{algorithmic}
\end{algorithm}

\newpage
\section{BystandAR}\label{appendix-bystandar}
As shown in Alg.~\ref{alg:bystandarwithgenericpetandevaluatar}, BystandAR maps each detected face to a 3D coordinate and spawns a corresponding virtual GameObject to maintain identity across frames. Tracking is achieved by checking whether the projected positions in subsequent frames overlap with existing GameObjects, using Unity’s \texttt{ Physics.OverlapBox}. The first overlapping match is taken as the continuing identity. While this approach works in simple scenarios, it fails in crowded or complex settings with multiple overlapping or crossing faces.

\begin{algorithm}[b!]
\caption{\textsc{BystandAR} control loop with \frameworkname's hooks. The highlighted block is replaced by our proposed modifications.}
\label{alg:bystandarwithgenericpetandevaluatar}
\begin{algorithmic}[1]
\State \textbf{Parameters:} sampling interval $N$, Mode $\in$ \{Baseline, EvaluatAR-Collect, EvaluatAR-Replay\}, QRDetector
\State FrameNum = 0
\State Detector $\gets$ FaceDetection
\While{True}
    \State 
    %\AlgMark{b-sense-start} 
    CameraFrame $\gets$ current camera frame;\; DepthFrame $\gets$ retrieve raw depth
    \State FrameNum$++$; \; $S_t \gets$ current eye gaze, voice input
    %\AlgMark{b-log1-start} 
    \If{Mode $\neq$ Baseline} 
          \State $H_t \gets$ HeadsetPose;\; $Q_t \gets$ QRDetector.Pose 
          \State EvaluatAR.setCurrentCameraCapture(CameraFrame)
          \If{Mode $==$ EvaluatAR-Replay} 
            \State $S_t \gets$ EvaluatAR.getCurrentFrameData() 
        \EndIf  %\AlgMark{b-log1-end}
    \EndIf %\AlgMark{b-sense-end}
    %\AlgMark{b-by-start} 
    \If {$S_t$ intersects with a face} 
        \State Increment eye/voice tracker for face
        \If {Eye-gaze/Voice history $>$ Threshold}
            \State Label face a subject
        \Else
            \State Label face becomes/remains bystander
        \EndIf
    \EndIf %\AlgMark{b-by-end}
    % \AlgMark{b-det-start} 
    \If {FrameNum $\geq N$} 
        \State FrameNum = 0
        \State faces $\gets$ Detector(CameraFrame)
        \For{each face $\in faces$}
            \State Transform 2D detection to 3D world space
            \colorbox{yellow}{\parbox{\dimexpr\linewidth-2\fboxsep}{%
            \If {face overlaps with an existing face}
                \State Replace current detection; reset TTL
            \Else
                \State Create new detection
            \EndIf %\AlgMark{b-det-end}
            }}
            %\AlgMark{b-obf-start} 
            \If {Application requesting sensor data} 
                \State Obscure bystander faces in frame
            \EndIf
        \EndFor
    \Else
        \If {Application requesting sensor data}
            \State Obscure bystander faces in frame
        \EndIf
    \EndIf %\AlgMark{b-obf-end}
    %\AlgMark{b-log2-start} 
    \If{Mode $\neq$ \textsc{Baseline}} 
        \If{Mode $==$ \textsc{EvaluatAR-Replay}}
            \State LogData $\gets (FrameNum, faces)$
        \Else
            \State LogData $\gets (FrameNum, Q_t, H_t, S_t)$
        \EndIf
        \State EvaluatAR.writeToLogsFile(LogData)
    \EndIf %\AlgMark{b-log2-end}
    \State Release frames to the application
\EndWhile
% \DrawTightBlock{blue}{b-sense-start}{b-sense-end}{3.4em}{-1em}{1ex}{2pt}{2pt}
% \DrawTightBlock{green!60!black}{b-log1-start}{b-log1-end}{24.2em}{-22.6em}{-1.8ex}{2pt}{2pt}
% \DrawTightBlock{orange}{b-by-start}{b-by-end}{3.5em}{-1em}{-1ex}{2pt}{2pt}
% \DrawTightBlock{black}{b-det-start}{b-det-end}{3.5em}{-1em}{-3ex}{2ex}{2ex}
% \DrawTightBlock{red}{b-obf-start}{b-obf-end}{3.5em}{-1em}{-3ex}{2ex}{2ex}
% \DrawTightBlock{green!60!black}{b-log2-start}{b-log2-end}{3.5em}{-1em}{-3ex}{2ex}{2ex}

\end{algorithmic}
\end{algorithm}

\textbf{Proposed Modifications.} We implemented and evaluated four modifications to BystandAR’s tracking pipeline. \emph{Naive Predicted Positioning (NPP)}, \emph{Kalman Predicted Positioning (KPP)}, \emph{Closest Depth (CD)}, and \emph{Hybrid method}. 

NPP extends the BystandAR algorithm by naively calculating and storing the predicted position of each detection in the next frame by assuming consistent motion (Algorithm~\ref{alg:nppandkpp}). The prediction is computed based on the difference in 3D position between the past frame and the current frame and assumes the same translation will occur. Whenever there is an overlap between two faces, this approach uses the Cartesian distance between the 3D position of the new detection with the predicted positions of the overlapping faces and chooses the one with the smallest distance. The KPP method instead uses a Kalman filter~\cite{bishop2001introduction} to calculate and store the predicted position of each bounding box (Algorithm~\ref{alg:nppandkpp}).

\begin{algorithm}[h]
\caption{Control loop for NPP and KPP modifications}
\label{alg:nppandkpp}
\begin{algorithmic}[1]
\setcounter{ALG@line}{20} 
    \If {face overlaps with an existing face}
        \State Compute smallest distance using predicted positions
        \State Replace current detection; reset TTL
    \Else
        \State Create new detection
    \EndIf
\end{algorithmic}
\end{algorithm}

The CD approach relies solely on the depth of the detections to resolve overlapping bounding boxes (Algorithm~\ref{alg:cd}). This approach relies on the accuracy of face detection and the mapping from 2D camera space to 3D world space provided by BystandAR~\cite{corbett2023bystandar}. Our implementation compares the Z-coordinate of overlapping bounding boxes to select the closest match.

\begin{algorithm}[h]
\caption{Control loop for CD modification}
\label{alg:cd}
\begin{algorithmic}[1]
\setcounter{ALG@line}{20} 
    \If {face overlaps with an existing face}
        \State Compute smallest distance in Z dimension
        \State Replace current detection; reset TTL
    \Else
        \State Create new detection
    \EndIf
\end{algorithmic}
\end{algorithm}

The Hybrid method fuses the KPP and CD predictions when computing the distance between the new inference and overlapping bounding boxes (Algorithm~\ref{alg:hybrid}). The distance value is the weighted sum of the measures used in KPP and CD (the weights used for KPP and CD are 0.2 and 0.8, respectively).

\begin{algorithm}[!t]
\caption{Control loop for Hybrid (KPP and CD) modification}
\label{alg:hybrid}
\begin{algorithmic}[1]
\setcounter{ALG@line}{20} 
    \If {face overlaps with an existing face}
        \State Compute weighted sum of KPP and CD distances
        \State Replace current detection; reset TTL
    \Else
        \State Create new detection
    \EndIf
\end{algorithmic}
\end{algorithm}

\section{Cardea-Inspired PET}\label{appendix-cardea}

As shown in Alg.~\ref{alg:cardeawithgenericpetandevaluatar}, the Cardea-inspired explicit PET enforces per-face visual obfuscation based on bystander intent gestures. Each frame, the PET runs face, hand, and gesture detection on the current camera frame. When at least one face is eligible for interaction, the PET associates detected gestures to faces using a proximity-based face–hand mapping in image space.

\begin{algorithm}[H]
\caption{Cardea-inspired PET's control loop with \frameworkname's hooks.}
\label{alg:cardeawithgenericpetandevaluatar}
\begin{algorithmic}[1]
\State \textbf{Parameters:} Gesture $\in$ \{OpenPalm, victory\}, Mode $\in$ \{Baseline, EvaluatAR-Collect, EvaluatAR-Replay\}, QRDetector
\State FaceDetector $\gets$ FaceDetection
\State HandPoseDetector $\gets$ HandPoseDetection
\While{True}
    \State 
    % \AlgMark{c-sense-start} 
    CameraFrame $\gets$ current camera frame %\AlgMark{c-sense-end}
        % \AlgMark{c-log1-start} 
        \If{Mode $\neq$ Baseline} 
          \State $H_t \gets$ HeadsetPose;\; $Q_t \gets$ QRDetector.Pose 
          \State EvaluatAR.setCurrentCameraCapture(CameraFrame) 
          % \AlgMark{c-log1-end}
    \EndIf %\AlgMark{c-sense-end}
    %\AlgMark{c-det-start} 
    \State faces $\gets$ FaceDetector(CameraFrame)
    \State handGestures $\gets$ HandPoseDetector(CameraFrame)
    % \If {activeFaces.Count $\neq$ 0}
    %         \State handGestures $\gets$ HandPoseDetector(CameraFrame) 
    %         \State handFaceMap $\gets$ (face, handGesture) proximity-based pairs
    % \EndIf %\AlgMark{c-det-end}
    \State handFaceMap $\gets$ (face, handGesture) proximity-based pairs
    %\AlgMark{c-obf-start} 
    \For{each mapping $\in handFaceMap$}
        %\AlgMark{c-by-start} 
        \If {$mapping[handGesture] ==$ openPalm}
            \State Apply blur for obfuscation
        \EndIf
        \If {$mapping[handGesture] ==$ victory}
            \State Remove obfuscation
        \EndIf
    \EndFor %\AlgMark{c-obf-end}
    %\AlgMark{c-log2-start} 
    \If{Mode $\neq$ \textsc{Baseline}} 
        \If{Mode $==$ \textsc{EvaluatAR-Replay}}
            \State LogData $\gets$ data of faces and their hand pose
        \Else
            \State LogData $\gets (FrameNum, Q_t, H_t)$
        \EndIf
        \State EvaluatAR.writeToLogsFile(LogData)
    \EndIf %\AlgMark{c-log2-end}
    \State Release frames to the application
\EndWhile
% \DrawTightBlock{blue}{c-sense-start}{c-sense-end}{3.4em}{-1em}{1ex}{-40pt}{2pt}
% \DrawTightBlock{green!60!black}{c-log1-start}{c-log1-end}{24.2em}{-22.6em}{-1.8ex}{15pt}{2pt}
% \DrawTightBlock{orange}{c-by-start}{c-by-end}{3.5em}{-1em}{-1ex}{2pt}{2pt}
% \DrawTightBlock{black}{c-det-start}{c-det-end}{3.5em}{-1em}{-3ex}{2ex}{2ex}
% \DrawTightBlock{red}{c-obf-start}{c-obf-end}{3.5em}{-1em}{-3ex}{2ex}{2ex}
% \DrawTightBlock{green!60!black}{c-log2-start}{c-log2-end}{3.5em}{-1em}{-3ex}{2ex}{2ex}

\end{algorithmic}
\end{algorithm}

\section{Case Study~2}
This appendix section consists of additional graphs for Experiment~2~(\S\ref{casestudy2-exp2}) for completeness. These plots represent the results of Cardea-inspired explicit PET under standardized replay of inputs on MQ3 and ML2 across two visual stimuli (Video~1: one bystander; Video~2: two bystanders) and two model stacks (high vs.\ low)

\begin{figure}[h]
    \centering
     \includegraphics[width=0.8\linewidth]{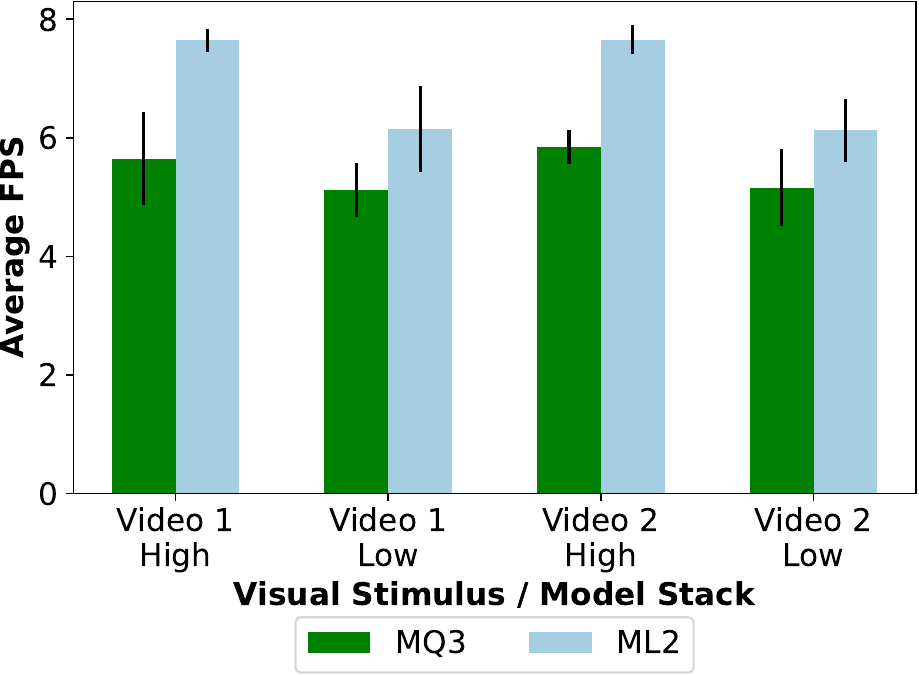}
     \caption{
     % Average FPS of the explicit PET across headsets, visual stimuli, and model stacks.
     Average FPS of the explicit PET across headsets, visual stimuli, and model stacks. ML2 consistently achieves higher FPS. Interestingly, the low-precision stack does not improve runtime performance.
     }
     \label{img:cs2_fps}
\end{figure}

\begin{figure}[h]
    \centering
     \includegraphics[width=0.8\linewidth]{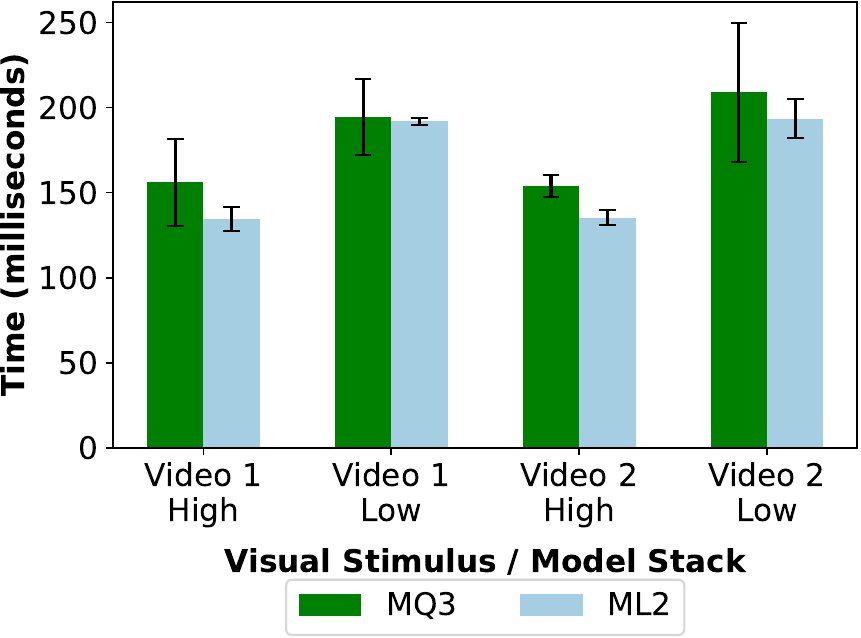}
     \caption{
     % Intent-to-enforcement processing time for the explicit PET across trials.
     Intent-to-enforcement processing time for the explicit PET across trials. Processing time varies with stimulus and model stack, but the low-precision stack does not offer much end-to-end latency benefit.
     }
     \label{img:cs2_latency_proxy}
\end{figure}

\end{document}

%% file: TextFiles/New_Introduction.tex
\section{Introduction}\label{introduction}

Augmented Reality (AR) headsets create immersive experiences by integrating multiple sensors, 
% including outward-facing cameras and depth sensors to understand the scene, microphones to capture audio, and biometric sensors to track eye gaze and body motion~\cite{Shepard_2022, Bacchus_2023, Lolambean_2023}. 
including outward-facing cameras, depth sensors, microphones, and biometric sensors (e.g., eye tracking)~\cite{Shepard_2022, Bacchus_2023, Lolambean_2023}. 
Advances in AR hardware~\cite{azuma2001recent, davis-2024, nikolaidis2022significant} have accelerated mainstream adoption, with market size projected to reach \$1.05 trillion by 2030, an estimated 29.7\% increase from 2025~\cite{grandview2026augmentedrealitymarket}. Increasing adoption of AR in industries like entertainment, healthcare, manufacturing, and education~\cite{michael-e-porter-and-james-e-heppelmann-2025, mekni2014augmented, chatzopoulos2017mobile, saballa-2022} exacerbates existing privacy issues 
% associated with the always-on sensing of AR headsets.
associated with its always-on sensing.

\begin{figure*}[t]
    \includegraphics[width=\textwidth]{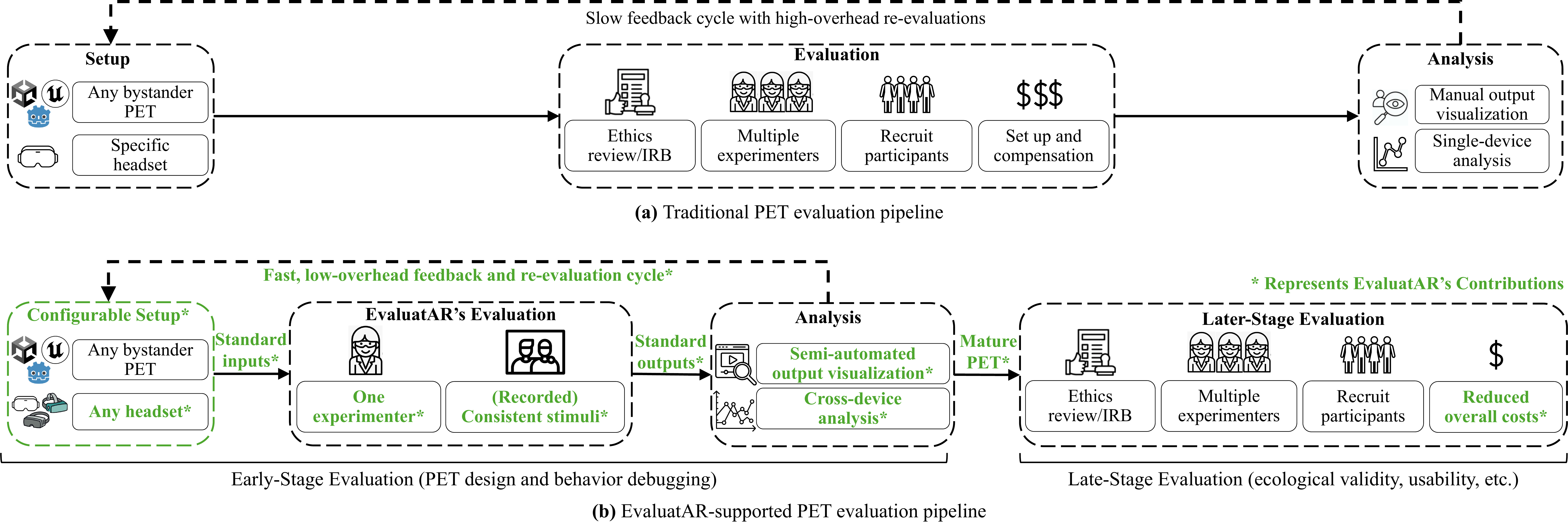}
    \caption{
    % (a) Traditional evaluations are often device-specific and rely on live user studies, requiring multiple experimenters and device-specific analysis. 
    % (b) \frameworkname\ enables controlled replication via replayable video-based visual stimuli and standardizes PET inputs/outputs that reduce evaluation overhead. 
    (a) Traditional evaluation is human-dependent and device-specific, coupling PET debugging and validation with live human-subject studies. This incurs substantial setup, recruitment/compensation, and experimenter overheads, while introducing confounds from variations in human behavior. 
    (b) \frameworkname\ introduces a low-overhead evaluation stage for controlled replication. This stage supports cross-device design validation, privacy-performance profiling, and rapid iteration before later-stage human-subject studies of mature PETs focused on ecological validity, user experience, and social acceptability.
    % \bo{if one compares traditional PET evaluation pipeline and late-stage evaluation after EvaluatAR in this new figure, would it give the impression that we only saved one dollar sign? we need to emphasize the repeated loop part}
    }
    \label{fig:teaser}
\end{figure*}

Always-on sensing poses a key privacy risk for nearby bystanders, especially when they are unaware of the sensing or do not consent to being recorded. Prior work shows that data captured by AR headsets can reveal sensitive personally identifiable information of bystanders, including facial identity, sex, race, age, and sexual preference~\cite{kroger2020does, liebling2014privacy, wenzlaff2016video, bye2019ethical, renaud2002measuring, borras2012depth, cheng20173d, giri2022emotion}. These risks are compounded by AR headset designs that often conceal sensors, making it difficult for bystanders to recognize when and how their privacy might be compromised~\cite{main2022evading, bukhari2025rethinking, sajid2025just}. This data may be shared without the consent of bystanders, resulting in undesired tracking or profiling~\cite{hill2022secretive, brennan2021federaldata}, and creating unease in bystanders regarding AR technology \cite{o2023privacy, gallardo2023speculative}. This unauthorized disclosure of bystanders' data is known as the bystander privacy protection (BPP) problem~\cite{corbett2023bystandar,corbett2023securing}, and visual sensing is a prominent driver of it in wearable technologies~\cite{lee2016information}.

% To address the BPP problem in AR, researchers have developed privacy-enhancing technologies (PETs) to protect bystanders' privacy in sensing pipelines (e.g., by detecting bystanders and applying privacy transformations such as obfuscation)~\cite{perez2018facepet, jana2013scanner, jung2014courteous, templeman2014placeavoider, corbett2023bystandar, yus2014faceblock}. In this paper, we focus on visual PETs as visual sensing is a prominent driver of privacy concerns for bystanders in wearable technologies~\cite{lee2016information}, making vision a practical first target for standardized evaluation. Visual bystander PETs are typically evaluated along three key dimensions: \emph{usability} (bystander perceptions such as comfort and acceptability of PET's outputs), \emph{privacy protection} (scenario-based protection metrics), and \emph{performance} (system responsiveness, often reported as frames per second (FPS)). For example, Corbett et al.~\cite{corbett2023bystandar} evaluate BystandAR via a user study with 16 participants to assess the usability of the system and report privacy protection metrics. However, their evaluation examines the system’s performance only on HoloLens~2. This example reflects a broader trend: PETs are typically evaluated under conditions that are difficult to reproduce, resulting in a \emph{device-specific} or \emph{limited} performance evaluation that restricts comparisons of PET performance and privacy-performance trade-offs across headsets.

To address BPP concerns in visual sensing pipelines in AR, researchers have developed privacy-enhancing technologies (PETs) to protect bystanders' privacy (e.g., detecting bystanders and applying privacy transformations such as obfuscation)~\cite{perez2018facepet, jana2013scanner, jung2014courteous, templeman2014placeavoider, corbett2023bystandar, yus2014faceblock}. 
These PETs are typically evaluated along three key dimensions: \emph{usability} (bystander perceptions such as comfort and acceptability of PET's outputs), \emph{privacy protection} (scenario-based protection metrics), and \emph{performance} (system responsiveness, often reported as frames per second (FPS)). 
For example, Corbett et al.~\cite{corbett2023bystandar} evaluate BystandAR through a user study with 16 participants to assess system usability through survey responses, report privacy protection metrics computed from the dataset, and evaluate its performance on the HoloLens~2 with dedicated experiments. % Similarly, Aditya et al.~\cite{aditya2016pic} evaluate I-Pic's usability through an online survey with 227 participants, report privacy protection metrics, and measure its system performance on the Nexus 5 phone.
This example exemplifies the existing PET evaluation pipeline that has several limitations.

% \textbf{Gap.} Comparing bystander PETs across AR headsets requires running PET configurations under identical inputs and visual stimuli, but the community lacks a standardized, reproducible way to do so. In practice, live, human-dependent evaluations make it difficult to control confounds such as head motion, bystander movement, and viewpoint changes, which introduce noise and obscure privacy-performance trade-offs. This limitation matters because performance is not merely a quality-of-experience concern, but it determines whether a PET can enforce privacy under real-time constraints (e.g., detecting and obfuscating bystanders without dropping frames or missing appearances).
\textbf{Gap.} The current PET evaluation pipeline is \emph{human-dependent}, i.e., it involves multiple experimenters and participants, introducing confounds due to variations in human behavior (bystander movement, viewpoint changes, etc.).
It also has \emph{high overheads}, i.e., it requires substantial investment in engineering effort, study design, participant recruitment and compensation, and relies on experimenter availability.
Moreover, system performance evaluation in the existing pipeline is \emph{device-specific} or \emph{limited}, i.e., performance is measured only on the device on which the PET is developed. % \textbf{These limitations make the existing PET evaluation pipeline difficult to reproduce, limiting our understanding of PETs' generalizability, particularly their cross-device performance and privacy-performance trade-offs.} 
\textbf{These limitations imply that the existing PET evaluation pipeline is difficult to reproduce, limiting our understanding of PETs' generalizability, particularly their cross-device performance and privacy-performance trade-offs.} 
% This implication matters because performance is not merely a quality-of-experience concern, but it determines whether a PET can enforce privacy under real-time constraints (e.g., detecting and obfuscating bystanders without dropping frames or missing appearances).

% We address this gap by introducing \frameworkname\ that standardizes the cross-device evaluation pipeline by abstracting device-specific interfaces and enforcing consistent PET inputs and outputs. Concretely, \frameworkname\ provides (1) a common PET input/output interface, (2) presentation of consistent visual stimuli, and (3) a record-replay workflow for sensor data consumed by a PET.

% To address this gap, we introduce \frameworkname\ at the early stage of the PET evaluation pipeline. The goal of evaluation at this stage is to test the PET’s design feasibility and robustness. \frameworkname\ solves the gap by standardizing the early-stage evaluation by abstracting device-specific interfaces and enforcing consistent PET inputs and outputs, enabling cross-device performance and privacy-performance trade-off evaluation of PETs.} Concretely, \frameworkname\ provides (1) a common PET input/output interface, (2) presentation of consistent visual stimuli, and (3) a record-replay workflow for sensor data consumed by a PET.
We address this gap by introducing \frameworkname, a framework that standardizes PETs' inputs (sensor data and visual stimuli) and outputs, to enable controlled cross-device evaluation of PETs under consistent evaluation conditions. We introduce \frameworkname\ at the early stage of the PET evaluation pipeline, focusing on testing PET's design feasibility and robustness. Hence, we focus our evaluation of PETs via \frameworkname\ on assessing their generalizability and to profile their performance and privacy-performance trade-offs, enabling rapid prototyping. In doing so, \frameworkname\ 
reduces the overheads associated with the existing PET evaluation pipeline by enabling the development of mature PET designs before later-stage, high-overhead human-subject studies that focus on testing PET’s ecological validity, user experience, and social acceptability.

% \textbf{Use Case.} \frameworkname\ targets early-stage PET development, where researchers need low-overhead, reproducible cross-device evaluation of PET's feasibility, robustness, and privacy-performance trade-offs before committing to resource-intensive human-subject studies. As shown in Fig.~\ref{fig:teaser}, traditional evaluation pipelines often have experiments requiring multiple individuals to recreate visual stimuli for device-specific analysis. \frameworkname\ lowers the personnel, time, and cost overhead of such evaluations through record and replay of PET inputs (sensor data and visual stimuli) and standardized output logging. This lets a single experimenter conduct comparable trials across headsets to diagnose failures and iterate on the PET before later-stage in-situ studies of mature systems.

\textbf{Use Case.} Consider a researcher who has developed a novel visual bystander PET and wants to evaluate it. Traditionally, as illustrated in Fig.~\ref{fig:teaser}a, this requires live human-subject studies, ethics review/IRB approval, participant recruitment and compensation, and device-specific data collection and analysis. Even with this overhead, the evaluation remains limited by the number of testable conditions, confounds from variations in human behavior, and difficulties assessing how the PET generalizes across headsets.

Using \frameworkname\ instead, they would perform a low-overhead early-stage evaluation centered on controlled replication. After selecting videos depicting relevant scenarios, they integrate their PET with \frameworkname\ through lightweight input/output hooks, collect scenario-synced sensor data, and replay these inputs across trials and headsets. This enables reproducible, low-overhead, cross-device evaluation for design validation, debugging, privacy-performance profiling, and rapid iteration on failure cases under controlled conditions. As illustrated in Fig.~\ref{fig:teaser}b, \frameworkname-supported pipeline helps researchers identify and address major PET design issues early on to produce mature PETs for later-stage evaluation,
focusing on ecological validity, user experience, and social acceptability. Doing so saves researchers costly redesigns and repeated iterations.

\textbf{Scope.} In this work, we present and evaluate \frameworkname\ using two PETs representing \textit{implicit} and \textit{explicit} design categories. Implicit PETs infer bystanders' privacy preferences through the sensed context, whereas explicit PETs require bystanders to communicate their privacy preferences. For our case studies, we choose BystandAR~\cite{corbett2023bystandar}, a state-of-the-art (SOTA) implicit PET that combines real-time computer vision with eye tracking, and a Cardea-inspired~\cite{shu2018cardea} gesture-driven explicit PET that uses active bystander input for privacy protection enforcement. Using these PETs, we present three case studies to validate \frameworkname’s capabilities: (1) controlled cross-headset replay of identical inputs to characterize device-specific privacy-performance trade-offs; (2) generalizability of \frameworkname's workflow across explicit and implicit PET design categories; and (3) rapid prototyping via a targeted failure-case study to diagnose privacy-relevant failures in PET outputs and validate proposed modifications, yielding an improved PET variant.

\textbf{Contributions.} Our work makes three main contributions.
\begin{enumerate}[topsep=0pt, itemsep=2pt, parsep=0pt]
    \item We present \frameworkname, a headset-agnostic, modular record-replay framework that enables reproducible evaluation of bystander PETs by standardizing their inputs (sensor data and visual stimuli) and logging comparable outputs, enabling low-overhead, cross-device PET evaluation.
    \item We validate \frameworkname\ on three widely used, commercially available AR headsets (Microsoft HoloLens~2, Magic Leap~2, and Meta Quest~3), demonstrating reproducible cross-device evaluation across sensing pipelines and on-device compute constraints while requiring lightweight integration effort.
    \item We show how \frameworkname\ supports rapid PET iteration by capturing and replaying privacy-relevant failures to compare PET variants under identical inputs, enabling a verified improvement to BystandAR, a state-of-the-art bystander PET.
\end{enumerate}

\begin{table}
  \caption{ 
  Summary of evaluation scope across surveyed visual PETs. \xmark: none; \emptycircle: limited; \filledcircle: extensive. 
  Prior work evaluates usability and privacy protection more extensively, but performance evaluation is often device-specific.
  % motivating the need for a reproducible cross-device framework such as EvaluatAR.
  }
  \label{tab:litreview}
  \resizebox{\columnwidth}{!}{
  \begin{tabular}{l|c|c|c}
    \toprule
    System Name & Usability & Privacy & Performance\\
    \midrule
    SnapMe~\cite{henne2013snapme} & \emptycircle & \xmark & \xmark \\
    FaceBlock~\cite{yus2014faceblock, pappachan2014semantic} & \xmark & \xmark & \xmark \\
    BlindSpot~\cite{patel2009blindspot} & \emptycircle & \emptycircle & \xmark \\
    PrivacyCamera~\cite{li2016privacycamera} & \xmark & \filledcircle & \emptycircle \\
    I-Pic~\cite{aditya2016pic} & \filledcircle & \filledcircle & \emptycircle \\
    Cardea~\cite{shu2018cardea} & \filledcircle & \filledcircle & \emptycircle \\
    Respectful Cameras~\cite{schiff2009respectful} & \xmark & \filledcircle & \emptycircle \\
    \begin{tabular}[c]{@{}l@{}}Your Privacy is in Your\\ Hand~\cite{shu2017your}\end{tabular} & \filledcircle & \filledcircle & \emptycircle \\ 
    PrivateEye~\cite{raval2016you} & \filledcircle & \filledcircle & \emptycircle \\
    WaveOff~\cite{raval2016you} & \filledcircle & \filledcircle & \emptycircle \\
    BystandAR~\cite{corbett2023bystandar} & \filledcircle & \filledcircle & \emptycircle \\
    MarkIt~\cite{raval2014markit} & \xmark & \emptycircle & \emptycircle \\
    Virtual Curtain~\cite{shrestha2023virtual} & \emptycircle & \emptycircle & \filledcircle \\
    PlaceAvoider~\cite{templeman2014placeavoider} & \xmark & \filledcircle & \filledcircle \\
    Courteous Glass~\cite{jung2014courteous} & \xmark & \filledcircle & \filledcircle \\
  \bottomrule
\end{tabular}
}
\end{table}

%% file: TextFiles/New_RelatedWork.tex
\section{Related Work}\label{relatedwork}

In this section, we review PETs for visual data~(\S\ref{relatedworks-petsforvisualdata}), their evaluation practices~(\S\ref{relatedworks-evaldimsofpetsforvisualdata}), and examine prior efforts to standardize their evaluation~(\S\ref{relatedworks-evalframeworksforpets}).

\subsection{PETs for Visual Data}\label{relatedworks-petsforvisualdata}

Developing PETs to protect sensitive information in visual sensing pipelines has been an active area of research. Corbett et al.~\cite{corbett2023bystandar} distinguish bystander PET designs into two categories:

% \textbf{Implicit PETs.} Implicit PETs infer privacy intent from sensor data and contextual cues without requiring bystander participation. They are often treated as SOTA in AR as they support AR's always-on sensing requirements without requiring bystander registration or cooperation. However, their reliance on sensor-rich, real-time inference makes them sensitive to device-specific sensing and compute pipelines, motivating the need for reproducible cross-device evaluation and debugging. An early example of an implicit PET is Courteous Glass~\cite{jung2014courteous} that uses on-device sensing (e.g., thermal imaging) to detect privacy-relevant situations. In AR headsets, BystandAR~\cite{corbett2023bystandar} represents a SOTA implicit bystander PET, leveraging eye tracking and social cues to provide real-time protection. 
\textbf{Implicit PETs.} These systems infer privacy intent from sensor data and contextual cues without requiring bystander participation. They rely on social cues (e.g., posing for a picture) or other contextual information (e.g., distance from the camera, eye gaze direction, emotion, head pose, position in the frame) to detect bystanders and obfuscate them~\cite{darling2019identification, hasan2020automatically}. An early example of an implicit PET is Courteous Glass~\cite{jung2014courteous} that uses on-device sensing (e.g., thermal imaging) to detect privacy-relevant situations.

% Implicit PETs are often treated as SOTA in AR as they support AR's always-on sensing requirements without requiring bystander registration or cooperation. However, their reliance on sensor-rich, real-time inference makes them sensitive to device-specific sensing and compute pipelines, motivating the need for reproducible cross-device evaluation and debugging.
% {\color{blue}
% For example, BystandAR~\cite{corbett2023bystandar} is a SOTA implicit PET for AR headsets that continuously processes the egocentric scene and combines real-time face detection with sensor-rich cues such as eye gaze and audio to distinguish subjects from bystanders. Since it runs the entire pipeline on-device, it is sensitive to device-specific sensing and compute constraints. Hence, we select BystandAR as our implicit case-study PET because it captures the core properties and challenges of this category in AR.
% }
The current SOTA implicit PET for AR headsets is BystandAR~\cite{corbett2023bystandar} as it supports AR's always-on sensing requirements without requiring bystander registration or cooperation. It runs continuously on-device and processes egocentric scene captures, performs real-time face detection on them, and distinguishes subjects and bystanders in the scene by leveraging sensor-rich cues such as eye gaze, audio, and spatial mapping. However, the system's reliance on sensor-rich, real-time inference makes it sensitive to device-specific sensing and compute pipelines, motivating the need to evaluate its cross-device feasibility, performance, and privacy-performance trade-offs. Hence, we select BystandAR as our implicit PET case-study as it captures the core properties and challenges of this category in AR.

% \textbf{Explicit PETs.} Explicit PETs enable bystanders (or users acting on their behalf) to directly specify what should be protected, including interaction-driven region specification or tagging~\cite{raval2016you,shrestha2023virtual,shu2017your}, preference registration and broadcast~\cite{templeman2014placeavoider,aditya2016pic,li2016privacycamera}, and visible markers or tags that signal consent~\cite{schiff2009respectful,patel2009blindspot}. Cardea~\cite{shu2018cardea} represents a hybrid explicit PET approach that combines contextual policy enforcement with an in-situ hand-gesture mechanism for opt-in/opt-out signaling at capture time.

\textbf{Explicit PETs.} These systems require the user or the bystander to perform some action to express their privacy preferences. While these systems provide stronger user/bystander control over what is protected, they also impose an interaction burden by requiring conscious participation, special equipment, or enrollment in advance. Prior approaches include preference registration and broadcast~\cite{aditya2016pic, li2016privacycamera, templeman2014placeavoider}, visible markers or tags~\cite{schiff2009respectful, patel2009blindspot}, and interaction-driven controls such as gestures, tagging, or region specification~\cite{shu2017your, raval2016you, shrestha2023virtual}.

A central challenge for explicit PETs is whether they can correctly detect an intent signal, associate it with the intended person, and update the protection state accordingly over time. Cardea~\cite{shu2018cardea} is a hybrid design in this space that combines contextual policy enforcement with direct privacy signaling at capture time via bystanders’ hand gestures. As it captures the core technical and interaction challenges of explicit PETs in AR, we use a Cardea-inspired gesture-driven PET as our explicit case study.

% \noindent We use these categories to structure our evaluation. Specifically, we instantiate one PET per design category: BystandAR as the SOTA implicit PET and a Cardea-inspired gesture-driven explicit PET. 
Together, these categories represent the design space of PETs that our framework must support. 
% Using one representative PET from each category allows our case studies to evaluate whether the same end-to-end workflow generalizes across the two dominant ways visual bystander privacy is operationalized in prior work.
Therefore, we instantiate one PET per category in our evaluation: BystandAR as the SOTA implicit PET and a Cardea-inspired gesture-driven explicit PET.

\vspace{-3mm}
\subsection{Evaluation of PETs for Visual Data}\label{relatedworks-evaldimsofpetsforvisualdata}

We surveyed the evaluations of 15 visual PETs. Our analysis identified three key evaluation dimensions that we use as an organizing lens rather than a strict taxonomy, since PETs differ in sensing pipelines, threat models, and deployment goals.

\textbf{Usability.} Usability refers to the perceptions of bystanders/users regarding their comfort and acceptability of a PET's outputs. Many systems, including Cardea~\cite{shu2018cardea}, WaveOff~\cite{raval2016you}, and BystandAR~\cite{corbett2023bystandar}, conduct in-person user studies to assess social acceptability, user comprehension, and interaction burden. Other systems, such as SnapMe~\cite{henne2013snapme}, use online surveys to collect perceptions at scale.

\textbf{Privacy Protection.} Privacy protection refers to the quantitative measures of bystander privacy protections. This dimension is commonly evaluated experimentally using annotated datasets or scenario-based tests across diverse environmental conditions. For example, PlaceAvoider~\cite{templeman2014placeavoider} and Courteous Glass~\cite{jung2014courteous} test privacy logic across settings (e.g., public vs.\ private spaces) and under different detection criteria, offering broader scenario coverage.

\textbf{Performance.} Performance refers to system responsiveness under resource and latency constraints, often reported as FPS. While many systems report FPS, CPU/GPU usage, or energy consumption~\cite{li2016privacycamera,aditya2016pic,raval2016you}, these measurements are frequently collected on a single device, often the platform for which the PET was designed. For instance, Android-based systems such as PrivacyCamera~\cite{li2016privacycamera} and WaveOff~\cite{raval2016you} were evaluated on a Nexus 5 smartphone, and BystandAR was evaluated solely on HoloLens2~\cite{corbett2023bystandar}. Moreover, existing evaluations are often live and human-dependent, making it difficult to reproduce the same 
% sensor-driven stimuli (e.g., viewpoint changes and motion patterns)\bo{it might be better to mention such examples of stimuli in the intro and Sec. 3, instead of here?} 
inputs (visual stimuli and sensor data) 
across trials. As a result, the literature provides limited support for \emph{reproducible, cross-device} comparison of PET performance and privacy-performance trade-offs.

In Table~\ref{tab:litreview}, we code prior work’s evaluation scope as none, limited, or extensive for each evaluation dimension. \emph{Extensive} indicates an in-person user study for usability, evaluation across multiple distinct environments/conditions or annotated datasets for privacy protection, and measurements on multiple devices for performance; \emph{limited} otherwise. Our survey suggests a consistent trend: usability and privacy protection are often evaluated with relatively extensive methodologies, whereas performance is often evaluated in a limited manner, as evaluation pipelines are difficult to replicate. Since real-time performance can directly influence whether privacy transformations are applied reliably in practice, this fragmentation motivates standardized, cross-device evaluation pipelines that can isolate design choices and expose privacy-performance trade-offs.
% , a gap \frameworkname\ targets.

\subsection{Evaluation Frameworks for AR Visual PETs}\label{relatedworks-evalframeworksforpets}

A small number of research efforts aim to standardize evaluation in AR and visual sensing systems, but these primarily emphasize user-centric usability and do not provide end-to-end support for evaluating bystander PETs in terms of privacy protection and real-time performance. For instance, Choong et al.~\cite{choong2022augmented} and D{\"u}nser \& Billinghurst~\cite{dunser2011evaluating} present structured usability evaluation approaches assessing cognitive load, interaction fidelity, and visual coherence for users. Archie~\cite{lehman2020archie} supports in-the-wild usability testing of AR systems, but it does not incorporate bystander privacy perspectives. Collectively, these approaches provide valuable usability insights but do not address evaluation settings where privacy is a dynamic system behavior involving bystanders and real-time sensing.

Erdélyi et al.~\cite{erdelyi2018privacy} move closer to our focus by proposing an objective framework to evaluate privacy protection filters for visual data through privacy-utility trade-offs. However, their framework assumes static, offline image datasets and evaluates filters in a post-processing context. This assumption does not align with real-time, multi-sensor AR PETs where latency, tracking stability, and device performance directly affect whether privacy transformations are enforced reliably. In addition, their framework focuses on visual data alone, whereas AR bystander PETs may depend on additional sensor context (e.g., gaze and spatial cues) and must operate under sensor-driven variability.

% Commercial AR SDKs and toolkits (e.g., ARCore~\cite{googleARDevelopers}, MRTK~\cite{mrtk3Overview}, Vuforia~\cite{vuforiaDeveloperPortal}) provide application-level utilities for capturing and replaying subsets of interaction and tracking data (e.g., head pose and controller/hand input), which can assist with debugging and reproducing \emph{application sessions} rather than standardized evaluation of bystander PETs across headsets. In particular, these SDK utilities do not provide (i) a standardized PET integration interface for supplying a PET with its required inputs and capturing its outputs in a consistent format across headsets, (ii) standardized visual stimuli and synchronized record-and-replay of the sensor data that a PET consumes (e.g., camera/depth alongside head pose and gaze), (iii) support for reproducible cross-headset comparisons under equivalent stimuli despite differences in sensing and runtime pipelines, or (iv) built-in runtime monitoring to measure PET system performance and privacy protection under controlled conditions. While a developer could engineer parts of this functionality on top of SDK utilities, doing so typically requires substantial, device-specific integration and instrumentation and still does not yield an end-to-end, standardized pipeline for reproducible cross-headset PET evaluation. As a result, evaluations built directly on SDK record/replay features tend to remain device-specific and difficult to reproduce and compare across headsets.
Commercial AR SDKs and toolkits (e.g., ARCore~\cite{googleARDevelopers}, MRTK~\cite{mrtk3Overview}, Vuforia~\cite{vuforiaDeveloperPortal}) provide platform-specific utilities for recording and replaying \emph{application sessions} (e.g., head pose and controller/hand input) to aid application debugging. However, these features do not standardize evaluation of bystander PETs across headsets: they do not support synchronized replay of the full sensor data streams a PET consumes (e.g., camera/depth with pose and gaze), nor do they provide reproducible visual stimuli and logging interfaces that enable cross-headset comparisons under the same inputs.

Overall, existing efforts show limited support for reproducible, cross-device evaluation of bystander PETs using standard stimuli.
% \frameworkname\ enables such headset-agnostic evaluation with comparable PET outputs and performance metrics, supporting their rapid prototyping.

%% file: TextFiles/New_Framework.tex
\vspace{-2.5mm}
\section{Framework Design}\label{frameworkdesign}

In this section, we introduce \frameworkname, a framework that enables low-overhead early stage bystander PET evaluation in a \emph{reproducible, headset-agnostic} manner. We summarize the common components of visual bystander PETs~(\S\ref{frameworkdesign-genericpet}), enumerate technical challenges that arise when designing for cross-device evaluation~(\S\ref{frameworkdesign-challenges}), and describe how \frameworkname\ operationalizes controlled evaluation through its record-replay workflow~(\S\ref{frameworkdesign-frameworkworkflow}). 

\begin{figure*}[t]
  \centering
  \includegraphics[width=0.9\linewidth]{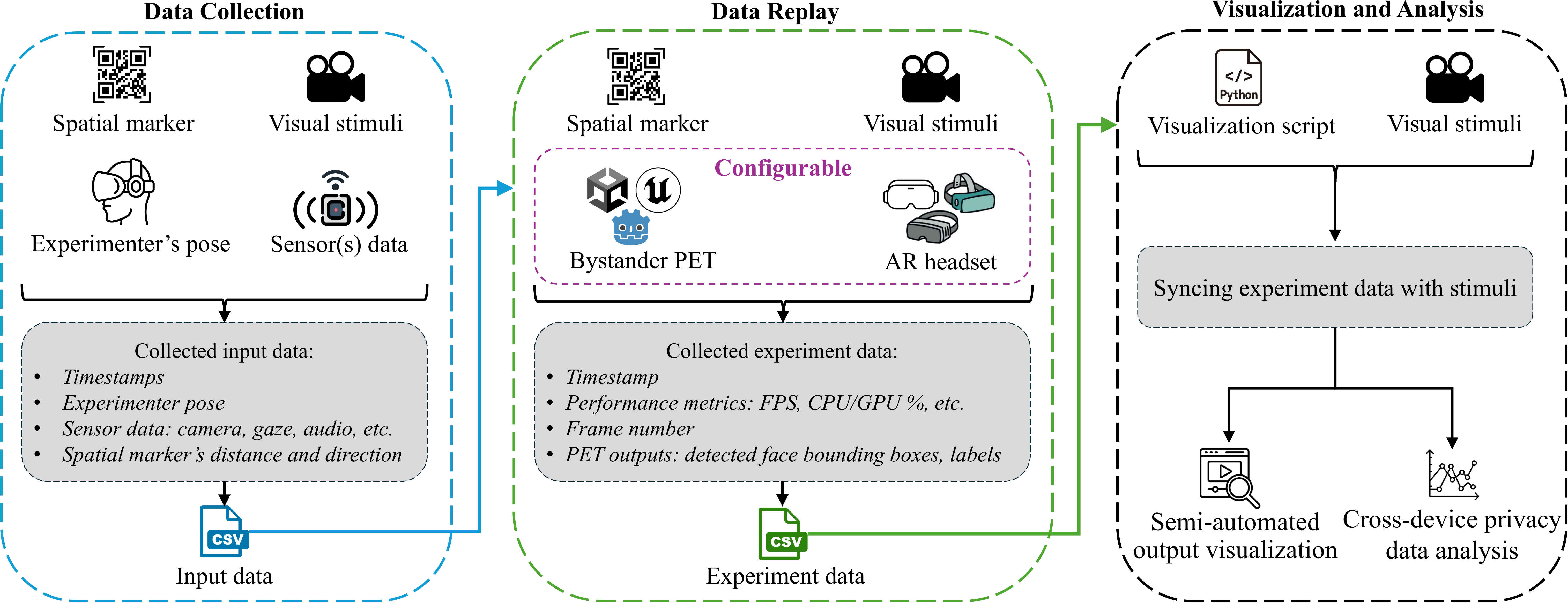}
  \caption{
  \frameworkname's three-stage workflow. \textbf{Data Collection} records PET inputs (sensor data and visual stimuli) and the experimenter’s pose relative to a spatial marker. \textbf{Data Replay} enables stimulus-synchronized replay of sensor data across headsets and trials, while logging performance metrics and PET outputs. \textbf{Visualization and Analysis} visualizes the logged experiment data, enabling cross-device performance and privacy-performance profiling of PETs under controlled conditions.
  }
  \Description{A graphic showing the three stages of the framework. It shows the inputs and outputs of each stage of the framework.}
  \label{img:frameworkvisualization}
\end{figure*}

\subsection{Components of Visual Bystander PETs}\label{frameworkdesign-genericpet}

% While bystander PETs vary in sensing modalities, detection models, and privacy transformations, most visual PETs follow a common runtime pattern consisting of the following components (Alg.~\ref{alg:genericpet}) that motivate our framework design:
While bystander PETs vary in sensing modalities, detection models, and privacy transformations, most visual PETs follow a common runtime pattern consisting of the following components (Alg.~\ref{alg:genericpet}):

\textbf{Sensing.} This component involves ingestion of egocentric sensor data streams~(Alg.~\ref{alg:genericpet}, lines~\ref{sensing1} and~\ref{sensing2}), including camera frames and optional sensor data that a PET may consume (e.g., eye gaze, depth, audio cues, or other context).

\textbf{Detection.} This component entails the detection of candidate bystanders~(Alg.~\ref{alg:genericpet}, line~\ref{detector}) in the scene (e.g., faces or full body segmentation). This process is typically performed through detectors that produce candidate regions (e.g., bounding boxes or masks).

\textbf{Bystander Decision Logic.} The goal of this component is to apply PET-specific logic to determine which candidates require protection, based on detections from the Detection stage, and optionally sensor data~(Alg.~\ref{alg:genericpet}, line~\ref{petbystanderlogic}). A possible default logic for a PET is to protect all candidate bystanders.

\textbf{Privacy Transformation.} This step applies PET-specific privacy transformation (e.g., obfuscation through blurring or pixel masking) to protected candidates in the visual output~(Alg.~\ref{alg:genericpet}, line~\ref{petobflogic}).

\begin{algorithm}
\caption{Generic visual bystander PET control loop.}
\label{alg:genericpet}
\begin{algorithmic}[1]
\State \textbf{Parameters:} Detector $\in FaceDetection, HumanSegmentation$
\While{True}
    \State CameraFrame $\gets$ current camera frame \label{sensing1}
    \State $S_t \gets$ sensor(s) data \label{sensing2}
    \State bboxes $\gets Detector(CameraFrame)$ \label{detector}
    \For{each $bbox \in bboxes$}
        \State isBystander $\gets$ PET\_BystanderLogic(bbox, $S_t$)\label{petbystanderlogic}
        \If{isBystander}
            % \State PET\_ObfuscationLogic(CameraFrame, bbox)\label{petobflogic}
            \State PET\_PrivacyTransformation(CameraFrame, bbox)\label{petobflogic}
            
        \EndIf
    \EndFor
\EndWhile
\end{algorithmic}
\end{algorithm}

% {\color{orange}
% These components operate on egocentric visual content and sensor inputs. Therefore, a reproducible cross-device evaluation framework must reliably replicate both the visual scene and the associated input data that a PET processes.
% }

\subsection{Main Challenges}\label{frameworkdesign-challenges}
\frameworkname\ enables low-overhead, reliable replication
of experimental conditions across headsets and trials for the early-stage
PET evaluation. The goals of this evaluation are to
debug PET behavior, validate design choices, and profile PET's performance and privacy-performance trade-offs before investing in high-overhead human-subject
studies. To support these goals, our initial design intuition for the framework was to capture scenarios as videos and replay them on a screen. Our rationale behind this design decision was that preserving these scenarios as video stimuli would eliminate the confounds inherent in existing setups requiring live human recreation (inconsistent head motion, bystander movement, or viewpoint changes, etc.). However, we observed that this approach had several technical challenges in practice. 

% Specifically, we observed the following technical challenges in the simple screen-based replay approach. 
\textbf{(C1) Viewpoint Inconsistencies} The generic PET components identified in~\S\ref{frameworkdesign-genericpet} establish that visual bystander PETs operate on egocentric visual content and sensor inputs. Thus, reliably reproducing the visual scene seen by the PET is critical. If the experimenter’s viewpoint changes across trials, PET may process different visual content even when the same video stimulus is replayed. Similarly, small variations in viewpoint can lead to significant differences in replayed sensor data. For example, eye gaze is cast from the wearer's head position, so small differences in the pose of the experimenter wearing a target headset can produce larger angular errors in gaze data for elements in the scene.
% Hence, we need a mechanism to reliably recreate experimenters' pose every time, or something along these lines?

\textbf{(C2) Repeated Device-Integration Effort}: APIs available for exposing sensor data to applications vary across AR platforms. This variation incurs non-trivial engineering costs, as implementing sensor data capture and replay requires device-specific implementations across PETs and target headsets.

\textbf{(C3) Synchronization Issue of PET Inputs}: AR headsets differ in processing speed, compute capabilities, and runtime behavior. These differences impact the synchronization of replayed sensor data with its associated stimulus. For example, a headset with superior hardware may replay sensor data faster than a headset with older hardware, leading to inconsistent data replay across headsets.

\textbf{(C4) Synchronization Issue of PET Outputs and Stimulus}: Variations in headset execution rates and stimulus playback cause PET outputs and visual stimuli to go out of synchronization. Desynchronization complicates the precise alignment required for reliable visualization and analysis for effective early-stage PET evaluation.

\vspace{-2.5mm}
\subsection{Framework Workflow}\label{frameworkdesign-frameworkworkflow}
% As described in~\S\ref{frameworkdesign-genericpet}, a PET’s behavior depends on the egocentric sensor inputs and the visual content within the headset's field of view (FoV). Hence, a reproducible, headset-agnostic evaluation framework must 
% (1) capture and replay sensor inputs under standardized visual stimuli, 
% (2) capture and replay the experimenter's pose (position and rotation) to preserve the viewpoint to control visual content processed by the PET across devices and experimental trials, and 
% (3) log PET outputs and performance measurements in a comparable format compatible across headsets. \frameworkname's end-to-end workflow achieves these goals in three stages: Data Collection, Data Replay, and Visualization and Analysis. 
The identified challenges motivate \frameworkname's controlled design. It uses video-based stimuli for standardized scene recreation rather than relying on live participants whose behavior could vary, introducing confounds. Similarly, \frameworkname\ uses a fixed-pose design for the experimenter to preserve viewpoint and to enable standardized replay of sensor data across headsets and trials. It also standardizes the PET inputs/outputs and adopts an elapsed time-based replay approach of sensor data through a common workflow to reduce device-specific engineering overheads and synchronization issues. \frameworkname\ operationalizes this design in three stages: Data Collection, Data Replay, and Visualization and Analysis.

\textbf{Data Collection.} 
The goal of this stage is to capture input data for consistent experiment replication in a headset-agnostic manner. This includes sensor data and visual stimuli (visual content presented to the wearer, such as video recordings depicting bystanders and scene activity). This stage aims to record these inputs in a device-independent representation so the data captured on one headset can be replayed on another, despite the differences in sensor APIs exposed by headsets.

To achieve this goal, \frameworkname\ records the experimenter’s pose (Alg.~\ref{alg:genericpetforevaluatar}, line~\ref{ln:pose}) over time relative to a spatial marker (e.g., a QR code) placed in the environment (Alg.~\ref{alg:genericpetforevaluatar}, line~\ref{ln:qr}) and any additional input sensor data streams that the target PET consumes (Alg.~\ref{alg:genericpetforevaluatar}, line~\ref{ln:sensor}), such as eye gaze, audio cues, or ultra-wideband and Bluetooth signals used for locating bystanders~\cite{aaraj2025blindspot}. The spatial marker allows the experimenter to maintain a consistent spatial reference between Data Collection and Data Replay to preserve the experimenter's viewpoint across trials, addressing \textbf{(C1)}.

To avoid confounding runtime performance measurements, the visual stimuli are presented externally rather than rendered by a separate in-headset application alongside the PET. All recorded input data is stored with elapsed timestamps and is written via the unified logging hook (Alg.~\ref{alg:genericpetforevaluatar}, line~\ref{ln:logwrite}), addressing \textbf{(C2),} in a CSV file that has extensive compatibility across headsets.

\textbf{Data Replay.} The goal of this stage is to replay the input data to evaluate a PET under standardized conditions across headsets and trials. During replay, the headset detects the visual marker to calculate its current marker-relative pose, then uses the recorded marker-relative start pose to guide the experimenter back to the correct starting position. A lightweight alignment mechanism (e.g., visual cues) assists the experimenter in reproducing the viewpoint. Marker detection is disabled once the viewpoints align to avoid additional performance overhead caused by the framework on a PET. The framework also captures a reference FoV image at this point (enabled by storing the current camera frame via the hook in Alg.~\ref{alg:genericpetforevaluatar}, line~\ref{ln:camera}) to support accurate overlays in the Visualization and Analysis stage.

After alignment, the visual stimulus is played externally while \frameworkname\ supplies the PET with input sensor data for replay (Alg.~\ref{alg:genericpetforevaluatar}, line~\ref{ln:replaydata}). However, as noted in \textbf{(C3)}, different headsets process frames at different rates, desynchronizing replayed input data and its associated stimulus. \frameworkname\ support elapsed time-based replay of inputs: at time $t$, the framework uses the most recent logged input data point with timestamp $\leq t$, addressing \textbf{(C3).} In parallel, \frameworkname\ logs per-frame data, including elapsed time, frame number, runtime performance metrics (e.g., FPS), and PET outputs (e.g., detected regions/labels or obfuscation states), to an experiment data CSV file via a unified logging hook (Alg.~\ref{alg:genericpetforevaluatar}, line~\ref{ln:logwrite}).

In our design, the experimenter coordinates the start of input data replay and visual stimulus playback to minimize additional instrumentation (e.g., casting/remote control) that could otherwise introduce confounds. 

\textbf{Visualization and Analysis.} This stage supports semi-automated inspection and analysis of a PET’s outputs 
% across headsets and experimental trials under standardized conditions 
by aligning per-frame experiment data logs with the recorded visual stimulus. The experiment data CSV contains per-frame experimental data logs corresponding to the visual stimuli presented during the experiments. \frameworkname\ synchronizes logged experimental data (e.g., detected regions/labels and other PET outputs) with the visual stimuli and renders them as an annotated .MP4 video. This visualization enables quantitative analysis (e.g., performance trends over time) and systematic inspection of privacy failure cases. When appropriate, it can also serve as input to downstream perception studies (e.g., bystander-facing evaluations of PET outputs via surveys).

To generate accurate overlays, \frameworkname\ uses the reference FoV image captured during Data Replay to identify the on-screen stimulus boundaries (top-left and bottom-right corners). These points define the mapping from camera-space coordinates in the experiment data logs (e.g., bounding boxes) to stimulus coordinate space so overlays align with the visual content presented during evaluation. Since headset execution and stimulus playback may differ in frame rate, as noted in \textbf{(C4)}, \frameworkname\ aligns experimental data logs to the visual stimulus frames using elapsed time and drops unmatched frames at the beginning of playback, producing a synchronized output suitable for both visualization and analysis, and addressing \textbf{(C4)}.

\begin{algorithm}
\caption{Generic visual bystander PET control loop with \frameworkname\ hooks for standardized record-replay and logging.}
\label{alg:genericpetforevaluatar}
\begin{algorithmic}[1]
\State \textbf{Parameters:} QRDetector, Mode $\in \{Collect, Replay\}$, Detector $\in FaceDetection, HumanSegmentation$
\State FrameNum = 0
\While{True}
    \State CameraFrame $\gets$ current camera frame
    \State FrameNum$++$
    \State $Q_t \gets QRDetector.Pose$ \label{ln:qr}
    \State $H_t \gets HeadsetPose$ \label{ln:pose}
    \State $EvaluatAR.setCurrentCameraCapture(CameraFrame)$ \label{ln:camera}
    \If{Mode $== Replay$}
        \State $S_t \gets EvaluatAR.getCurrentFrameData()$ \label{ln:replaydata}
    \Else
        \State $S_t \gets$ sensor(s) data \label{ln:sensor}
    \EndIf
    \State bboxes $\gets Detector(CameraFrame)$
    \For{each $bbox \in bboxes$}
        \State isBystander $\gets$ PET\_BystanderLogic(bbox, $S_t$)
        \If{isBystander}
            \State PET\_PrivacyTransformation(CameraFrame, bbox)
        \EndIf
    \EndFor
    \If{Mode $== Replay$}
        \State LogData $\gets (FrameNum, bboxes)$ 
    \Else
        \State LogData $\gets (FrameNum, Q_t, H_t, S_t)$ 
    \EndIf
    \State $EvaluatAR.writeToLogsFile(LogData)$ \label{ln:logwrite}
\EndWhile
\end{algorithmic}
\end{algorithm}

%% file: TextFiles/New_FrameworkImplementation.tex
\vspace{-5mm}
\section{Framework Implementation}\label{frameworkdesign-frameworkimplementation}

\begin{figure*}
  \centering
  \begin{subfigure}{0.35\linewidth}
    \centering
    \includegraphics[width=\linewidth]{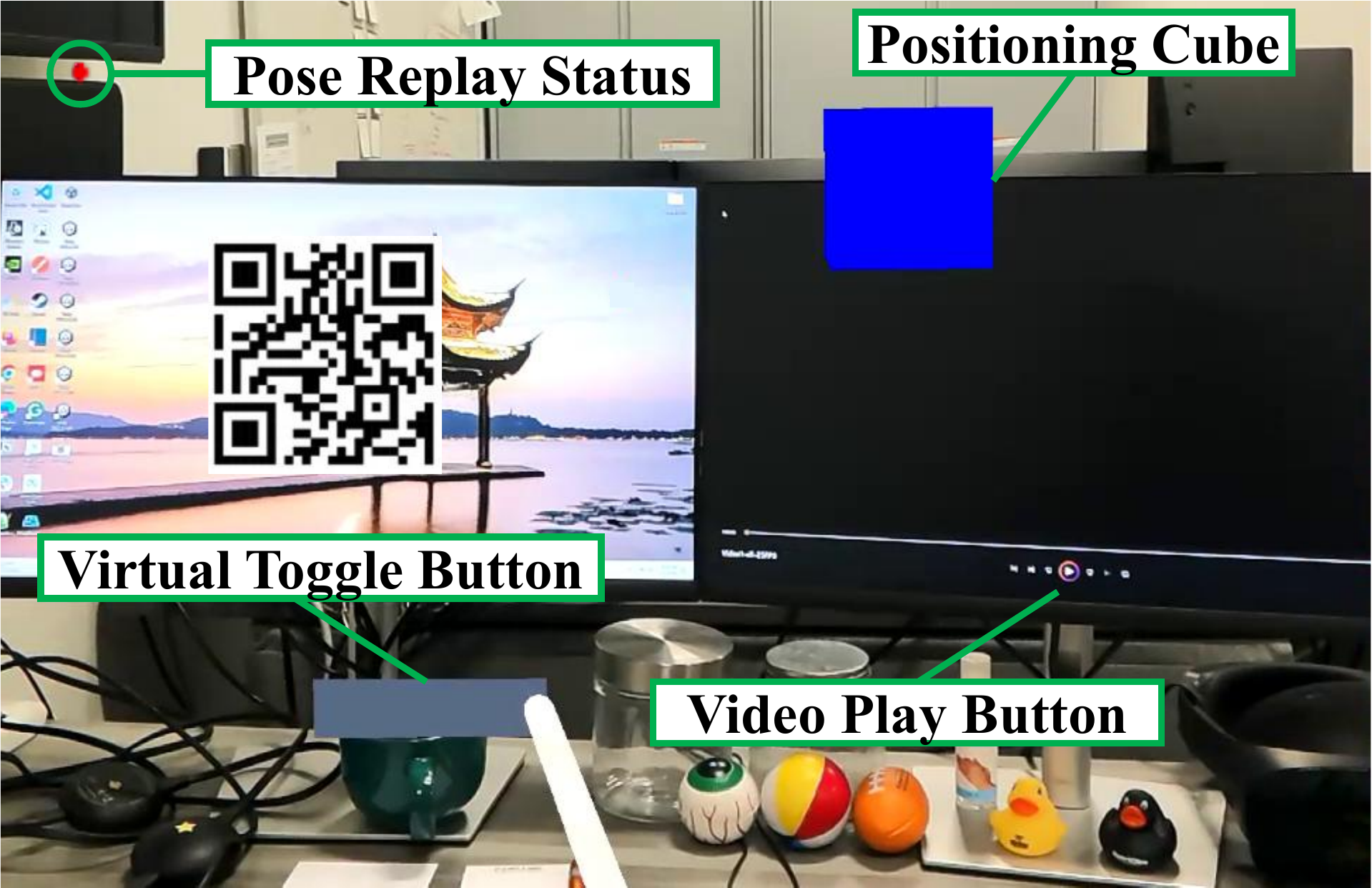}
    \caption{Before alignment}
    \label{fig:setup-before}
  \end{subfigure}
  \hspace{0.2cm}
  \begin{subfigure}{0.35\linewidth}
    \centering
    \includegraphics[width=\linewidth]{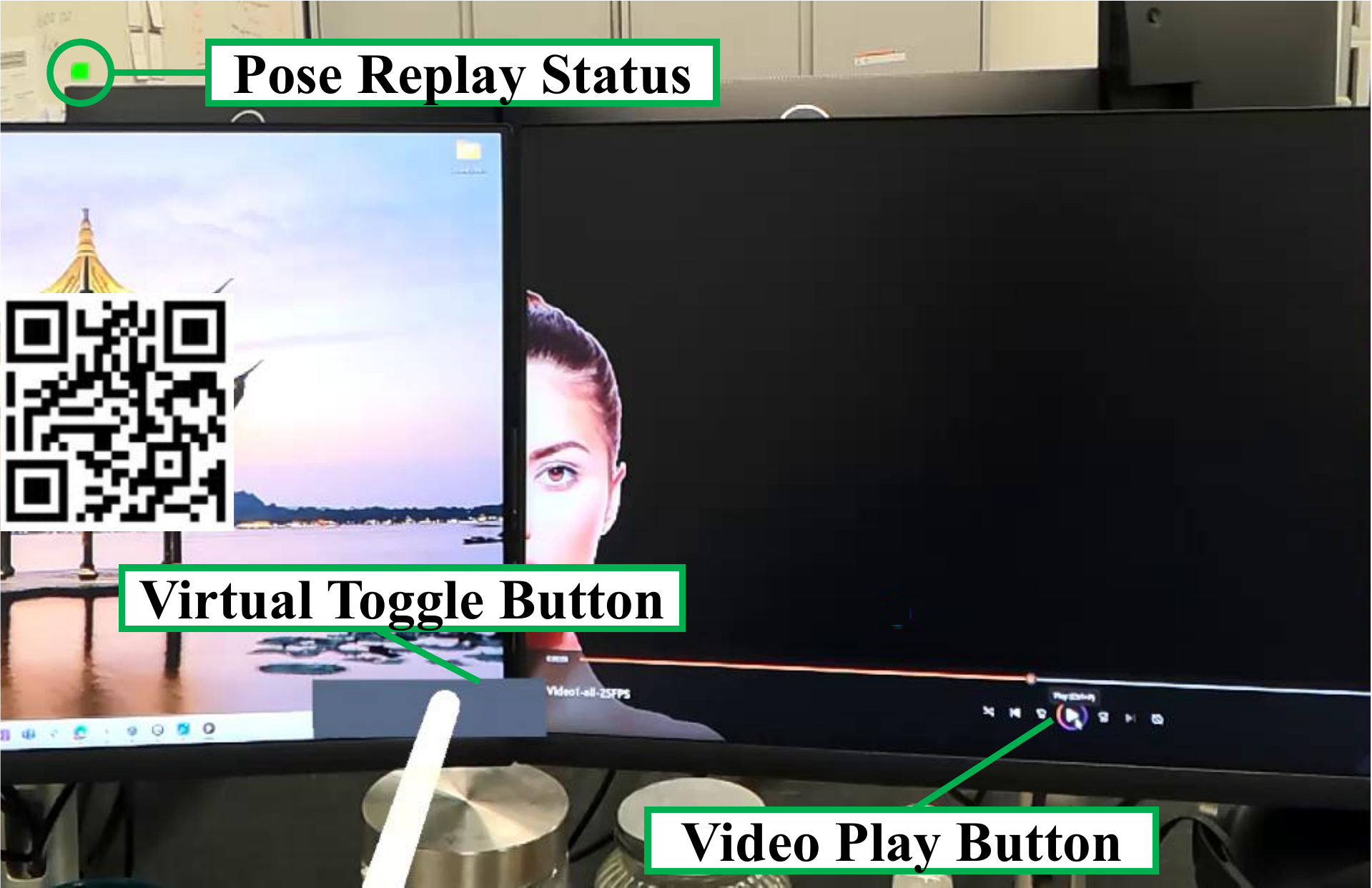}
    \caption{After alignment}
    \label{fig:setup-after}
  \end{subfigure}

  \caption{
  % \frameworkname's implementation. The QR marker and positioning cube facilitate the experimenter's viewpoint alignment across experimental trials. Pose Replay Status provides alignment feedback (turns from red to green when aligned), and the virtual toggle controls trial start/stop.
  \frameworkname's user interface before and after experimenter's viewpoint alignment across headsets and/or trials. The QR marker, positioning cube, and pose replay status indicator (turns from red to green when aligned) help guide the experimenter to the recorded start pose. Once aligned, the experimenter can start the trial via the virtual toggle button.
  }
  \label{img:experimentalsetup}
\end{figure*}

We implemented \frameworkname\ in Unity (2022.3.12f1) and made it available at 
% \textit{
% https://anonymous.4open.science/r/EvaluatAR-4842
\textit{https://github.com/SIMSB-99/EvaluatAR.git}. Our implementation uses a QR code as a shared physical marker, a virtual toggle button to start/stop logging, and a pose alignment guiding mechanism using a virtual positioning cube and a pose-status indicator (Fig.~\ref{img:experimentalsetup}). The algorithm of our implementation can be found in Appendix~\ref{appendix-evaluataralgo}.

\textbf{Data Collection.} \frameworkname\ logs one time-synchronized entry per frame in this mode until data recording is stopped by the experimenter. Each entry includes the timestamp, elapsed time since the start of data recording, frame number, the experimenter's pose, and the headset-to-marker relative pose (encoded as a 3D vector from the headset to the detected QR code).

\textbf{Data Replay.} Before running a PET with this mode of \frameworkname, the input data CSV created after Data Collection mode is added to the on-device storage of the target headset. When the PET runs, \frameworkname\ reconstructs the trial’s starting pose from the stored relative pose, and renders a virtual positioning cube at that location. The experimenter aligns their head pose with this cube until the pose-status indicator confirms alignment (Fig.~\ref{img:experimentalsetup}), after which \frameworkname\ disables QR detection for the remainder of the trial. \frameworkname\ also captures an RGB camera frame to support later mapping between camera space and the on-screen stimulus during Visualization and Analysis. The experimenter then begins the trial by pressing the toggle button and playing the visual stimulus video simultaneously.

\textbf{Visualization and Analysis.} We implement this stage as an offline Python script that processes collected experiment data, aligns the recorded headset FoV to the stimulus screen (annotated by the experimenter), and generates time-synchronized overlays of PET outputs 
% and performance metrics 
on the stimulus for inspection and analysis.

\noindent We instantiate this implementation across implicit and explicit PET designs and multiple headsets in our case studies~(\S\ref{casestudies}). To integrate a PET, we (1) expose the required inputs through lightweight getter functions in the PET’s main Unity script and (2) attach the \frameworkname\ script to a Unity GameObject to pass inputs and PET outputs into the framework’s logging interface. As our supported headsets already implement the remaining workflow infrastructure (as described above), adding a PET mainly requires integrating \frameworkname's input/output interfaces and validating the intended functionality. Based on the development time it took to integrate the MQ3 headset and Cardea-inspired explicit PET, we estimate 2-6 hours (one developer) for integration, plus 1-2 hours of validation per supported headset, totaling 3-8 hours to integrate and execute a new PET across all three headsets. These are informed engineering estimates rather than a controlled time study.

%% file: TextFiles/New_CaseStudies.tex
\vspace{-3mm}
\section{Case Studies}\label{casestudies}

% We design three case studies with the following goals that demonstrate \frameworkname's ability to:
% \begin{itemize}
%     \item \textbf{G1:} Enable reproducible, headset-agnostic evaluation of a bystander PET.
%     \item \textbf{G2:} Support evaluation of implicit and explicit PET design categories using the same end-to-end workflow.
%     \item \textbf{G3:} Enable PET iteration by comparing proposed modifications for privacy-relevant failures exhibited by a PET under standardized replay of edge cases.
% \end{itemize}
We conduct a case-study-based evaluation to validate whether \frameworkname\ supports representative early-stage PET evaluation goals. These goals include investigating PET's feasibility across headsets, robustness under controlled conditions, profiling its performance and privacy-performance trade-offs across supported devices, and debugging via replayable failure cases to support rapid iteration. This technical validation serves as a precursor to standalone privacy and usability evaluations, and high-overhead human-subject studies centered on ecological validity, user experience, and social acceptability, which are beyond the scope of this work.

We design three case studies with the following goals that demonstrate \frameworkname's ability to support early-stage PET evaluation:
\begin{itemize}
    \item \textbf{G1:} Enabling reproducible, headset-agnostic evaluation of a bystander PET’s performance.
    \item \textbf{G2:} Supporting evaluation of privacy-performance trade-offs across implicit and explicit PET design categories using the same end-to-end workflow.
    \item \textbf{G3:} Enabling PET design validation and iterative debugging by comparing proposed modifications for privacy-relevant failures exhibited by a PET under standardized replay of edge cases.
\end{itemize}

We use three commercially available headsets in our case studies: HoloLens~2 (HL2), Magic Leap~2 (ML2), and Meta Quest~3 (MQ3). These headsets span the two dominant AR display systems: HL2 and ML2 are optical see-through (OST) headsets, where users view the real world directly through transparent displays, whereas MQ3 is a video see-through (VST) headset, where the real world is mediated through a camera passthrough pipeline. These headsets also represent distinct on-device compute profiles: ML2 uses a quad-core Zen2 CPU (4×3.92 GHz, 8 threads), MQ3 uses an octa-core Kryo CPU (1×3.19 GHz, 4×2.8 GHz, 3×2.0 GHz), and HL2 uses an older octa-core Kryo385 CPU (4×2.96 GHz, 4×1.8 GHz)~\cite{VRcompare}. Together, these headsets represent different sensing pipelines and compute capabilities, allowing our case studies to demonstrate that \frameworkname\ supports reproducible early-stage evaluation across a broad range of modern AR headsets.

To reflect the breadth of bystander PET designs in prior work, we evaluate one implicit sensor-rich PET (BystandAR~\cite{corbett2023bystandar}) and one explicit gesture-driven PET (inspired by Cardea~\cite{shu2018cardea} and adapted for AR headsets). Using these two PETs lets our case studies demonstrate that the same end-to-end workflow can be applied across the two dominant PET design categories. In addition to the per-frame trial logs~(\S\ref{frameworkdesign-frameworkworkflow}), we record PET-specific outputs needed for analysis. For BystandAR, we log per-frame face detections and associated PET state (e.g., IDs, regions/predictions, and subject/bystander labels), and for the Cardea-inspired PET, we log per-module processing times, recognized gestures with associated face IDs, and per-face obfuscation state over time. We validate experimental trial integrity using the logged pose alignment status and expected replay progression; invalid trials are discarded and re-run.

Across our case studies, we draw the video-based visual stimuli from three sources: (1) commercially licensed stock footage, (2) author-recorded clips 
, and (3) synthetic sequences for controlled edge cases.
% For Case Study~1 and~2~(\S\ref{casestudy-performance} and~\S\ref{casestudy-explicit}), we use recordings that provide representative scene diversity for our case-study goals (e.g., different scenarios, participant appearances, motion regimes, crowding, camera-to-subject distance, and occlusion patterns). For Case Study~3~(\S\ref{casestudy-iteration}), we use synthetic videos where we require precise control over overlap/crossing trajectories to reproduce failure modes. Our use of these videos is an evaluation-scope choice to enable controlled replication and privacy-performance profiling. However, EvaluatAR’s record-replay pipeline is independent of how stimuli are sourced, so future instantiations can swap in more ecologically valid recordings without changing the workflow. Additionally, as some stimuli are commercially licensed, we do not redistribute raw stock footage. Instead, we have made \frameworkname's implementation available so others can reproduce the workflow using their own stimulus set.
For Case Studys~1 and~2, we use recordings that capture diverse real-world
scenarios relevant to bystander privacy, including coworkers in a
laboratory, newscasting and press conference settings, friends
cooking in a kitchen, people walking on a busy street, and virtual meeting-like scenes. These scenarios vary in motion, crowding, camera-to-subject distance, and occlusion patterns, allowing us to test PET behavior under representative scene diversity while retaining controlled replay. For Case Study 3, we use synthetic videos where precise overlap and crossing trajectories are required to reproduce privacy-relevant failure modes. This use of recorded and synthetic stimuli is an evaluation-scope choice to enable early-stage PET evaluation. Future instantiations can swap in more ecologically valid recordings without changing \frameworkname's workflow. Additionally, as some stimuli are commercially licensed, we do not redistribute raw stock footage. Instead, we have made \frameworkname's implementation available so others can reproduce the workflow using their own stimulus set.

\begin{figure*}[!t]
    \centering
    \makebox[\textwidth]{

        \begin{subfigure}{0.26\textwidth}
            \centering
            \includegraphics[width=\linewidth]{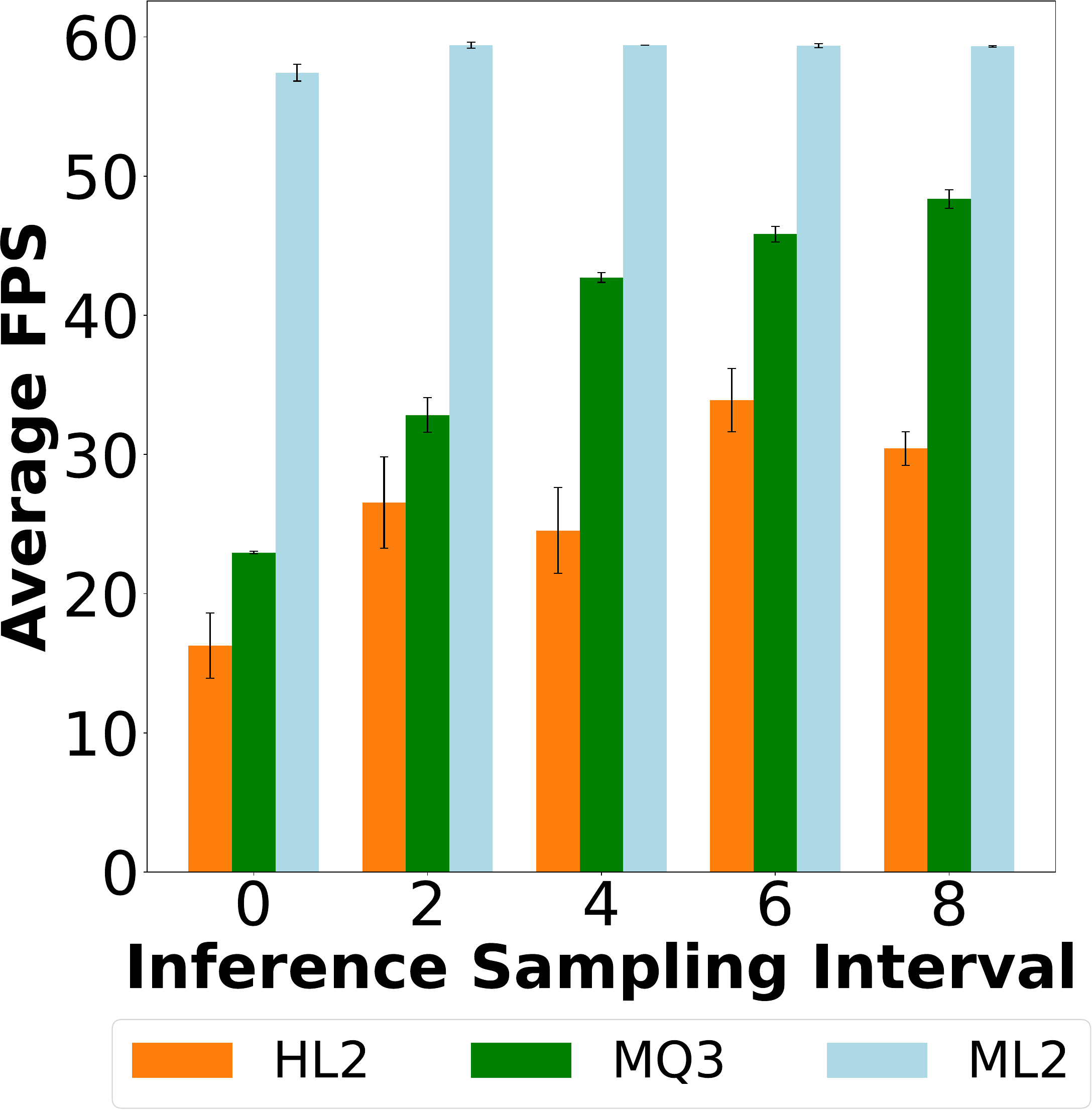}
            \caption{Minimal head motion (static)}
            \label{img:exp1-avgfpsvid1}
        \end{subfigure}%
        
        \hspace{0.001\textwidth}
        
        \begin{subfigure}{0.26\textwidth}
            \centering
            \includegraphics[width=\linewidth]{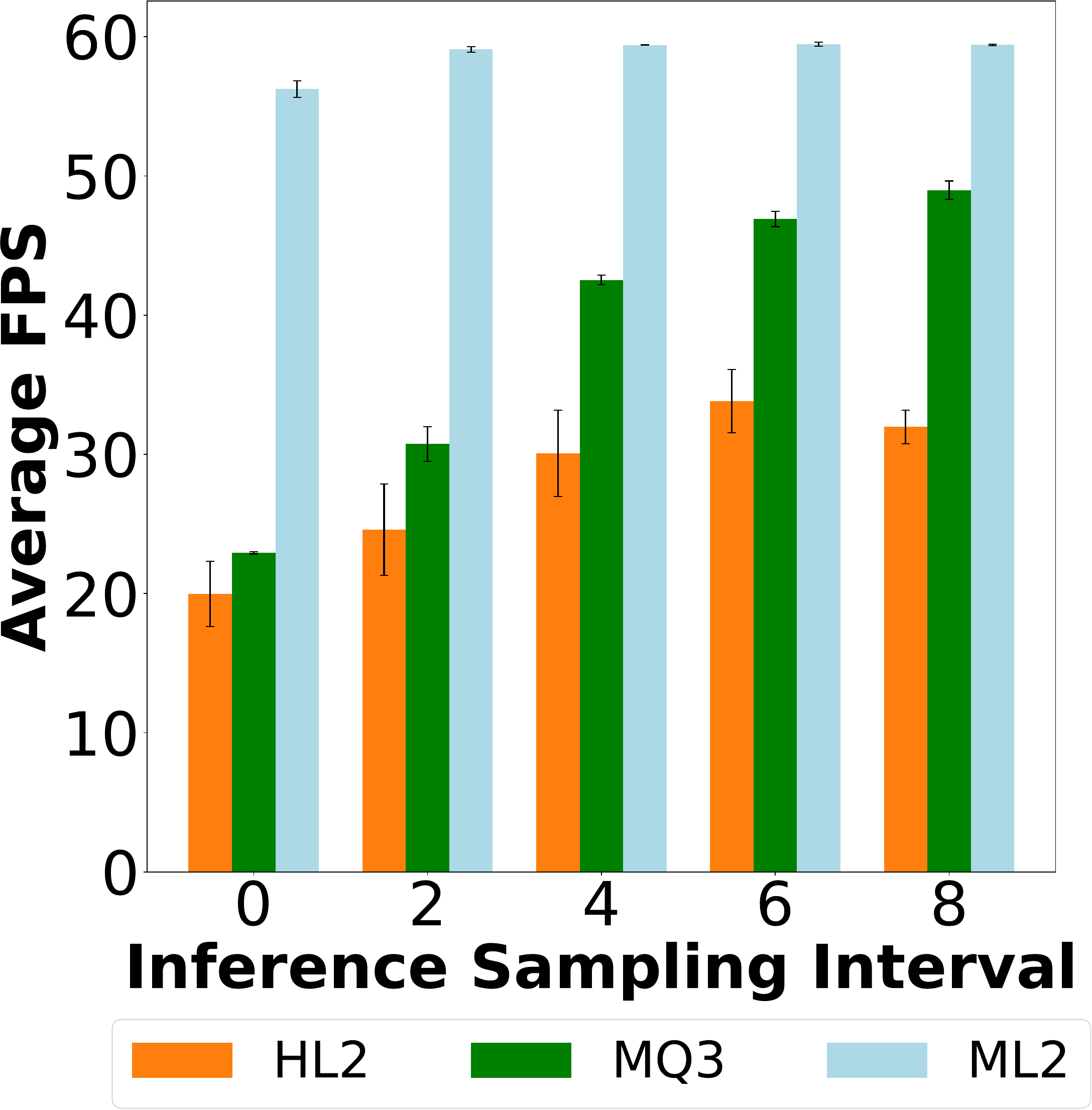}
            \caption{Gradual head motion (slow)}
            \label{img:exp1-avgfpsvid2}
        \end{subfigure}%
        
        \hspace{0.001\textwidth}
        
        \begin{subfigure}{0.26\textwidth}
            \centering
            \includegraphics[width=\linewidth]{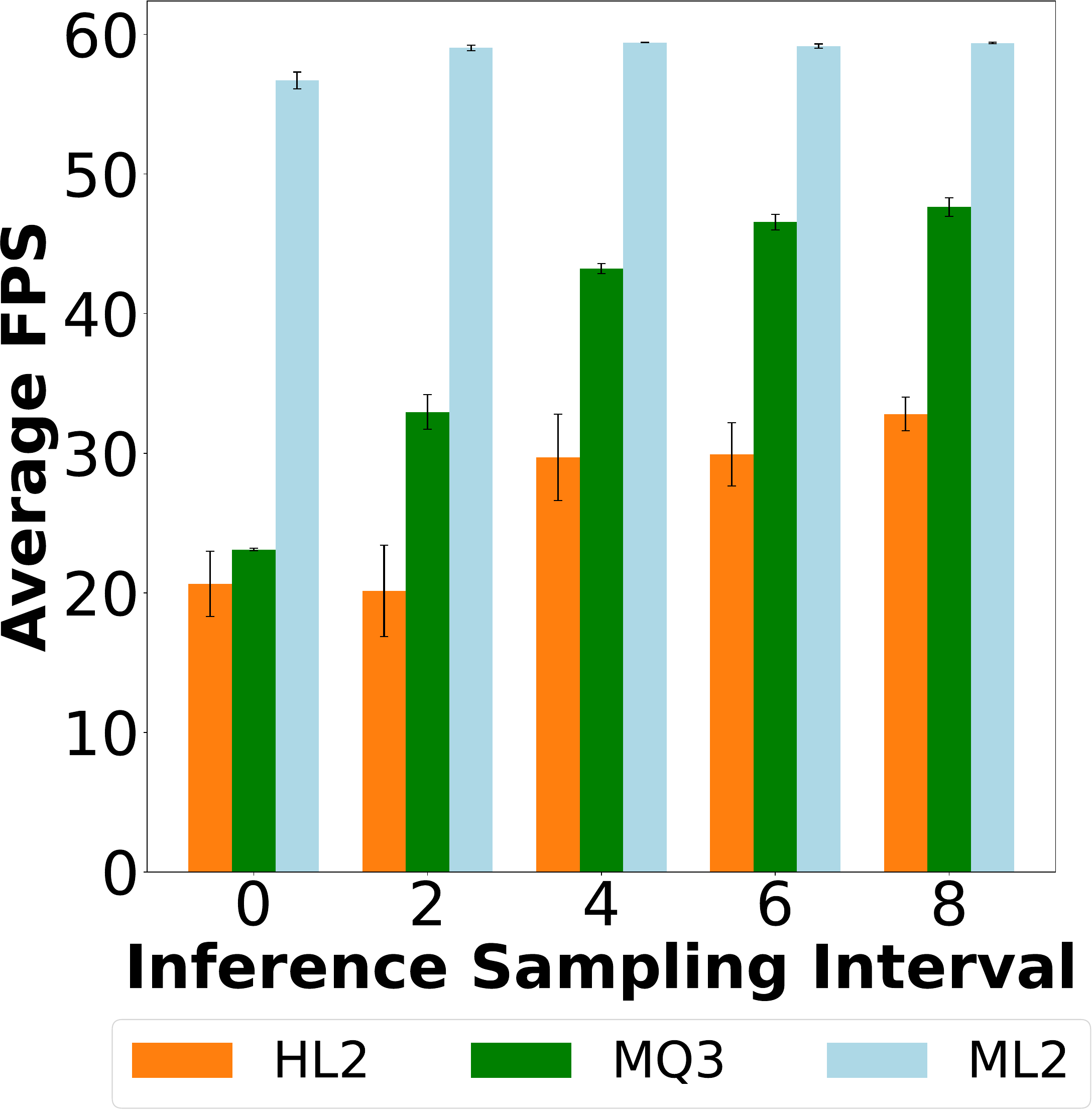}
            \caption{Rapid head motion (fast)}
            \label{img:exp1-avgfpsvid3}
        \end{subfigure}%
        
        \hspace{0.001\textwidth} 
        
        \begin{subfigure}{0.26\textwidth}
            \centering
            \includegraphics[width=\linewidth]{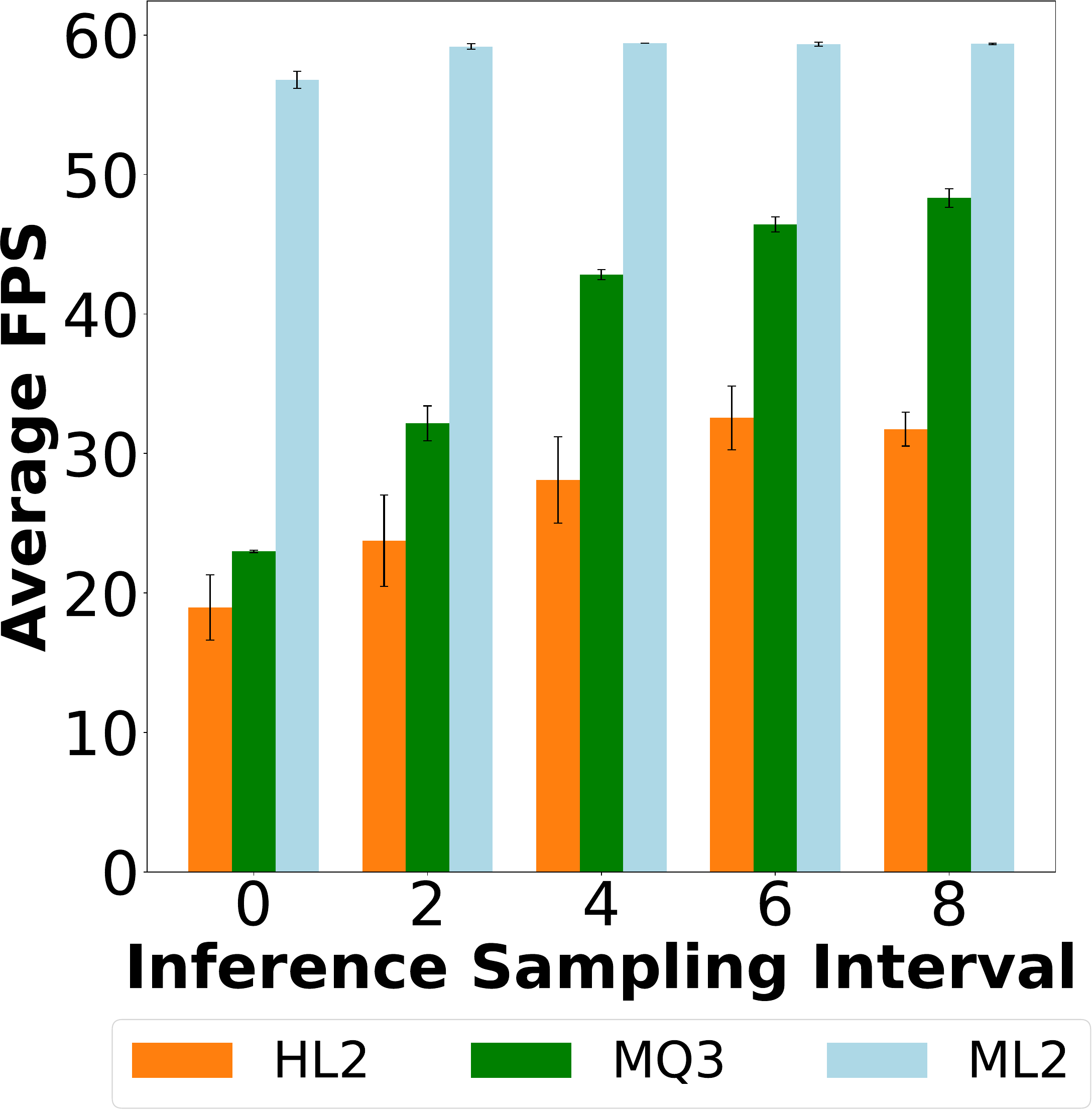}
            \caption{All videos of varying motions}
            \label{img:exp1-avgfpsperskipframevalue}
        \end{subfigure}
    }
    
    \caption{
    % Comparison of average FPS for different inference sampling intervals across the three headsets for the three videos with varying motion, and their averages from Experiment~1A. The value of 0 for the inference sampling interval represents inference on every frame.
    Average FPS across inference sampling intervals (0 represents per-frame inference) for the three headsets across static, slow, fast, and averaged motion conditions in Experiment~1A. Larger sampling intervals increase PET's performance. ML2 consistently achieves the highest performance, followed by MQ3 and HL2.
    }
    \label{img:exp1A}
\end{figure*}

\vspace{-3mm}
\subsection{Case Study~1}\label{casestudy-performance}

This case study is designed to achieve \textbf{G1} by demonstrating that \frameworkname\ can produce reproducible and comparable performance measurements for the same bystander PET when replayed under identical inputs on different headsets. 

We instantiate \frameworkname\ with BystandAR~\cite{corbett2023bystandar}, a SOTA implicit bystander PET for AR headsets, evaluated only on HL2 originally. BystandAR works by continuously processing the egocentric camera stream to detect candidate bystanders (face detection) and combines these detections with additional sensor data (notably eye gaze, audio cues, and depth data) to implement its bystander decision logic. Intuitively, BystandAR uses gaze-derived social cues to distinguish subjects (people the wearer is engaged with) from bystanders, and then applies a privacy transformation (blacking out associated pixel regions) to bystanders in the output stream. To remain real-time on device, BystandAR exposes an \emph{inference sampling interval} that skips a specified number of frames between inferences, directly impacting compute cost and real-time responsiveness. The detailed algorithm with \frameworkname\ integration can be found in Appendix~\ref{appendix-bystandar}.

We select BystandAR as the target PET for this case study because it represents a realistic, on-device bystander protection workload with two properties that make cross-device evaluation well-defined and interpretable. First, BystandAR has a published configuration knob (i.e., inference sampling interval) that systematically trades inference frequency for compute cost, allowing us to sweep a PET configuration range while holding the PET’s core logic constant. Second, BystandAR’s per-frame workload increases with the number of candidate bystanders visible in the scene: more people in view increase the number of detections that must be processed and maintained over time (e.g., tracking/association and per-person privacy transformations). These properties allow us to examine two common and practical sources of performance variation under standardized replay: (1) a PET-side configuration knob~(\S\ref{casestudy1-exp1a}) and (2) visual workload variation via candidate load~(\S\ref{casestudy1-exp1b}).

\subsubsection{Experiment 1A: PET configuration knob}\label{casestudy1-exp1a}

We design this experiment with inference sampling interval and headset as independent variables (IVs). We evaluate five levels of inference sampling interval IV: $\textit{interval} \in \{0,1,2,4,8\}$, where $\textit{interval}=0$ runs inference every frame and larger values skip more frames. We cap interval values at 8 to match the configuration range evaluated in the original BystandAR paper so that our evaluation remains within BystandAR’s published operating range, as our goal in this case study is to measure cross-device performance under standardized replay, not to redesign BystandAR's logic. The three headsets (HL2, MQ3, and ML2) represent the three levels of headset IV. The dependent variable (DV) is performance, measured via FPS.

\textbf{Procedure.} We use three videos drawn from commercially licensed stock footage as standardized visual stimuli for all trials. Each video contains a single visible person whose motion varies: minimal head motion (static), gradual head motion (slow), and rapid head motion (fast). Each headset is tested for all five $interval$ values across all three videos, resulting in a total of 75 trials.

\textbf{Results.} Across all three videos of varying motion (Fig.~\ref{img:exp1-avgfpsvid1},~\ref{img:exp1-avgfpsvid2}, and~\ref{img:exp1-avgfpsvid3}), increasing the sampling interval yields a clear increase in FPS on every headset. This pattern is consistent across the static, slow, and fast videos, indicating that the impact of the configuration knob is stable across different motions under standardized replay. 
% Importantly for \textbf{G1}, this monotonic trend is observed under reproducible conditions enabled by \frameworkname\ for standardized replay of inputs, meaning that changes in FPS reflect the intended configuration change 
% rather than differences in user behavior or scene re-enactment.
% rather than being confounded by variations in experimental conditions and procedure.

The results also expose a stable cross-headset ordering under the same replayed inputs and configurations: ML2 sustains the highest FPS across the tested interval values, MQ3 achieves moderate FPS, and HL2 remains substantially slower even at larger interval values. Since \frameworkname\ replays the same inputs and visual stimuli on each headset, these differences can be attributed to device/runtime constraints, showcasing that \frameworkname\ enables reproducible cross-headset comparison of the same PET.

Finally, we use these results to select a per-headset operating configuration for subsequent experiments. We define the best interval for a headset as the smallest interval value at which the FPS reaches the performance plateau (i.e., further increases in interval yield only marginal gains), thereby avoiding unnecessary skipping of inference while still operating near the performance knee. Using this criterion, we select $\textit{interval}=8$ for HL2, $\textit{interval}=4$ for MQ3, and $\textit{interval}=2$ for ML2. 
% Overall, Experiment~1A demonstrates how \frameworkname\ supports reproducible cross-headset performance profiling of a fixed PET workload under standardized replay, as required by \textbf{G1}.

\subsubsection{Experiment 1B: Candidate bystander load}\label{casestudy1-exp1b}

This experiment evaluates how PET performance scales as the number of candidate bystanders in the visual stimulus increases under standardized replay, as this scaling behavior was not examined in the original BystandAR evaluation. We designed this experiment with the candidate bystander load and headset as IVs. We operationalize candidate load as the number of bystanders visible in the scene and evaluate nine load levels: $\textit{load} \in \{1,2,3,4,5,7,8,10,12\}$. The three headsets represent the three levels of headset IV. The DV is performance, measured via FPS.

\textbf{Procedure.} We use one segmented stimulus video to standardize the visual workload across trials. The video is composed of nine sequential scenes that depict the candidate load levels, separated by blank screens for segmentation. Lower-load scenes are extracted from stock footage, while the highest-load scenes are composited from these licensed clips to preserve facial detail when many bystanders are present. We replay this video for each headset, resulting in a total of three trials.

\begin{figure}
 \centering
 \includegraphics[width=0.8\columnwidth]{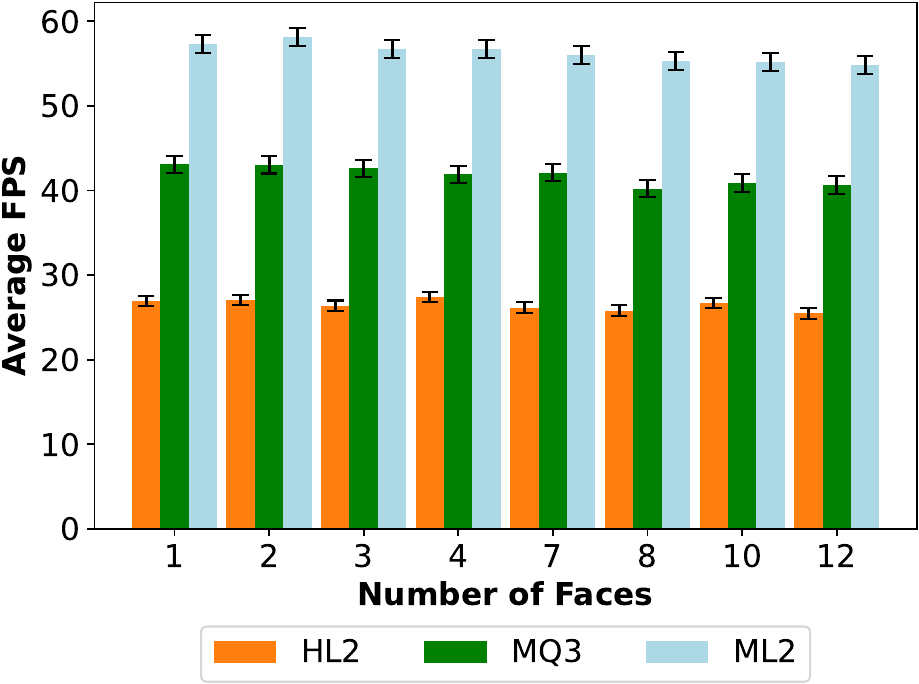}
 \caption{
 % Comparison of average FPS for HL2, MQ3, and ML2 across varying candidate bystander load tested in Experiment~1B.
 Average FPS across headsets against varying candidate bystander load in Experiment~1B. Headsets differ in baseline performance (ML2 performs the best, followed by MQ3 and HL2) and how PET's performance degrades (MQ3 shows the steepest decline, followed by ML2 and HL2).
 }
 \label{img:exp2-fpsvsgtforbothheadsets}
\end{figure}

\textbf{Results.} % \bo{I am not sure if the difference in the decline is significant enough to be worth a discussion point here: }
Fig.~\ref{img:exp2-fpsvsgtforbothheadsets} further reveals that headsets differ not only in their baseline FPS, but also in how quickly they lose performance as load increases. HL2 starts from the lowest FPS and shows comparatively limited additional decline with increasing load, suggesting it reaches a constrained operating point early. MQ3 exhibits the steepest decline as load increases, indicating tighter compute margins under higher candidate loads. 
% Fig.~\ref{img:exp2-fpsvsgtforbothheadsets} shows a clear decline in FPS as candidate load increases on all three headsets. This behavior is expected because higher-load scenes contain more people to be processed within the frame budget (e.g., more detections to maintain and more per-person privacy transformations), increasing the per-frame workload even though the replayed stimulus is identical across headsets. 

% ML2 maintains the highest FPS across the evaluated range and degrades more gradually, suggesting it sustains acceptable performance over a wider workload range under the same replayed inputs. Together, these results illustrate why cross-headset profiling must consider both baseline performance and scaling behavior: a headset can show a smaller drop with load yet still operate at an unusably low FPS, whereas a higher-capability headset can tolerate larger increases in candidate load while remaining usable.

\noindent Taken together, these experiments demonstrate that \frameworkname\ enables reproducible, cross-headset performance evaluation of the same PET under standardized replay. Since each experiment replays identical inputs and visual stimuli, the observed trends reflect the intended experimental variations for each experiment rather than being confounded by differences in experimenter behavior, scene content, or experimental procedure across trials. Under these reproducible conditions, \frameworkname\ yields consistent within-headset trends when sweeping a PET configuration knob and scaling visual workload, while also revealing stable cross-headset differences under the same conditions. This combination of repeatable trends within a headset and directly comparable measurements across headsets provides empirical support for \textbf{G1} and illustrates how \frameworkname\ can be used to identify feasible operating configurations for a PET on each device under controlled, replayed inputs.

\vspace{-3mm}
\subsection{Case Study~2}\label{casestudy-explicit}

Case Study~1 demonstrated \frameworkname’s standardized workflow on an implicit PET. We design Case Study~2 to advance \textbf{G2} by showing that the same end-to-end workflow can also be used to evaluate privacy-performance trade-offs of an explicit PET design while using PET-specific measures appropriate for its trigger mechanism.

We instantiate \frameworkname\ with a Cardea-inspired~\cite{shu2018cardea} gesture-driven explicit PET adapted for AR headsets. At runtime, the PET processes the egocentric camera stream and runs a multi-stage on-device perception pipeline (face detection, hand detection, and gesture recognition). It then associates a detected hand to a face using spatial proximity and applies a privacy transformation (Gaussian blur) based on the recognized gesture. It also tracks bystanders across frames by matching face detections with the highest overlap. Following Cardea’s definition, all people in view are treated as bystanders: by default, bystanders are not obfuscated; a bystander can explicitly request protection (opt-in) via an Open Palm gesture and revoke protection (opt-out) via a Victory (forming a V with index and middle finger) gesture. The detailed algorithm with \frameworkname\ integration can be found in Appendix~\ref{appendix-cardea}.

We select this PET because it represents an interaction-driven design of bystander PETs where privacy enforcement depends not only on per-frame performance, but also on whether the system correctly translates observable intent into the intended privacy state over time. This makes it a complementary workload to Case Study~1: under standardized replay, \frameworkname\ can capture both performance and intent-to-enforcement behavior using the same workflow, enabling a direct demonstration of generalizability across PET design categories (\textbf{G2}).

\subsubsection{Experiment 2: Model stack configuration and candidate load}\label{casestudy2-exp2}

We design this experiment to stress two practical factors that shape explicit PET behavior under real-time constraints: (1) the compute-accuracy trade-off in the PET’s perception pipeline and (2) the difficulty of maintaining correct hand-face association when multiple bystanders are present. We therefore use model stack configuration
% \matt{recommend a rewrite here to make this easier to understand for non-ML folks. I.e., "model inference variables i.e., model stack configuration" or something} 
, candidate bystander load, and headset as IVs. Model stack has two levels: $\textit{stack} \in \{high, low\}$ where \textit{high} represents the high-precision stack that uses the base versions of the face, hand, and gesture models, and \textit{low} represents the low-precision stack that replaces them with INT8 (quantized) variants. Candidate load has two levels, $\textit{load} \in \{1,2\}$, corresponding to one-bystander versus two-bystander scenes. We evaluate on two headsets (ML2 and MQ3). We attempted to include HL2, but the gesture pipeline exhibited unstable runtime behavior (very low FPS and frequent crashes), preventing consistent Data Replay runs; we therefore treat HL2 as a feasibility boundary for this particular PET and focus the comparisons across ML2 and MQ3.

% We report two classes of DVs. First, we report performance via FPS and per-module processing times (milliseconds) for face, hand, and gesture stages (plus the obfuscation step). Second, we report the PET's intent-to-enforcement behavior over time via (1) reliability: whether each scripted opt-in/opt-out intent event within the recorded visual stimuli yields the correct per-face obfuscation state, and (2) responsiveness via intent-frame pipeline cost proxy: for each intent event, we compute the total processing time by summing the module times required to produce and apply the obfuscation decision (e.g., face, hand, gesture, and blur).
We report two classes of DVs. First, we report performance via per-frame processing time (ms) and its module-level breakdown (face, hand, gesture, blur). Second, we measure intent-to-enforcement behavior via (1) reliability: whether each scripted opt-in/opt-out intent event within the recorded visual stimuli yields the correct per-face obfuscation state, and (2) responsiveness via intent-frame pipeline cost proxy by summing the per-module times required to produce and apply the obfuscation decision.

\textbf{Procedure.} The visual stimuli for this experiment were recorded by the authors. We recorded two standardized replay videos with scripted intent events so that bystander intent is controlled and repeatable under identical visual stimuli. Video~1 contains a single bystander performing repeated Open Palm (opt-in) and Victory (opt-out) gestures, interleaved with non-gesture periods to test stability. Video~2 increases candidate load by including two bystanders, while only one performs the scripted gesture sequence to stress hand-face association under higher candidate load. For each headset, we replay each video under both model stack configurations and repeat each condition three times, resulting in 24 total trials.

\begin{figure}
 \centering
 \includegraphics[width=0.88\columnwidth]{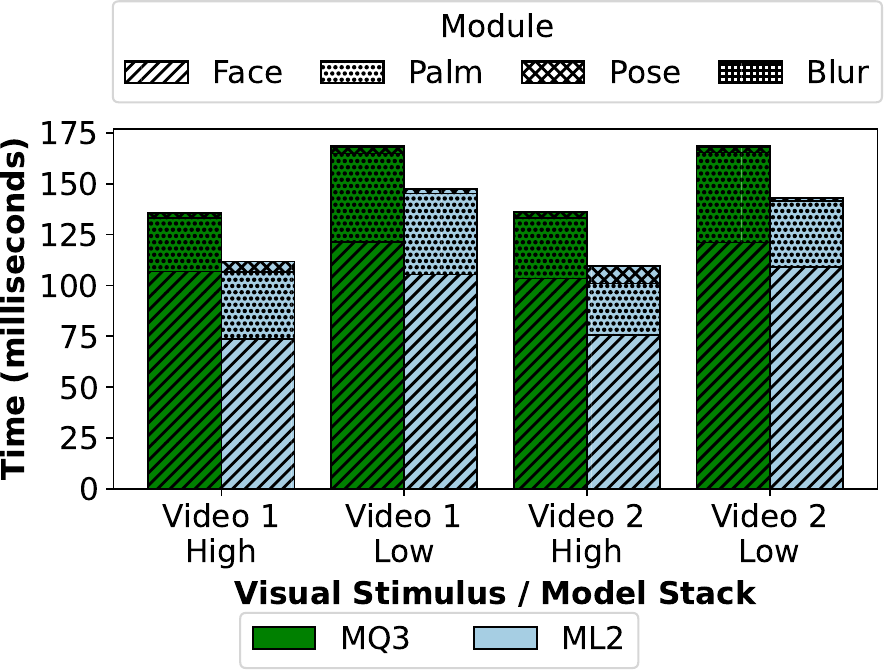}
 \caption{
 % Module-level processing time for the explicit PET pipeline, showing the contribution of face, palm, gesture, and blur modules across trials.
 Module-level processing time for the explicit PET pipeline. The face detector takes the most time, dominating the overall pipeline cost. Interestingly, the low-precision stack does not reduce end-to-end processing time.
 }
 \vspace{-4mm}
 \label{img:cs2_stage_ms}
\end{figure}

\textbf{Results.} Under identical replayed inputs, ML2 consistently sustains lower total per-frame processing time (higher FPS) than MQ3 (lower FPS) across both videos and both model stacks (Fig.~\ref{img:cs2_stage_ms}). This mirrors the stable cross-headset ordering observed in Case Study~1~(\S\ref{casestudy-performance}), where ML2 maintained the highest FPS across headsets. However, the absolute operating performance remains much lower (ML2: 7 FPS, MQ3: 5.5 FPS) because the explicit PET executes a multi-stage detection pipeline on-device (face, hand, gesture, and obfuscation) on every frame. Moreover, the configurations with higher per-frame cost (MQ3 and the low-precision stack) also incur higher intent-to-enforcement processing cost, meaning the PET reacts more slowly to opt-in/opt-out events under these settings.

\begin{figure}
 \centering
 \includegraphics[width=0.77\columnwidth]{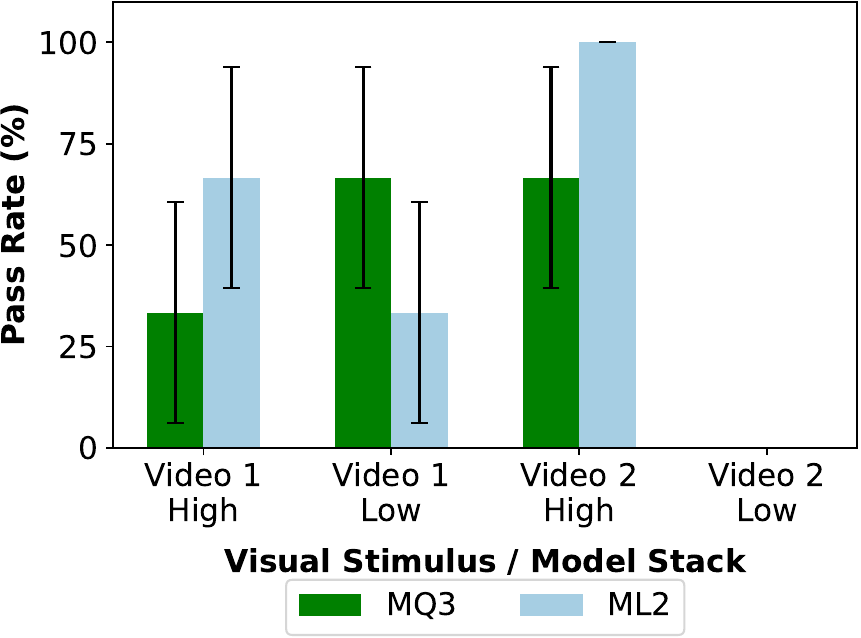}
 \caption{
 % Intent-to-enforcement correctness for the explicit PET (Experiment~2), reported as pass rate (\%), across visual stimuli (Video~1 vs. Video~2) and model stack configurations (high vs. low) on MQ3 and ML2.
 Intent-to-enforcement correctness of explicit PET. Correctness is more stable on ML2 and degrades across headsets under demanding configurations, especially with two bystanders (Video~2) and the low-precision stack (Low), where the Video~2 + Low condition reaches 0\% pass rate.
 % \bo{no bar plot for Video 2 Low?}\ibrahim{the pass rate for that is 0, that's why there is no bar for it}
 }
 \vspace{-4mm}
 \label{img:cs2_toggle_cond}
\end{figure}

Notably, the low-precision stack does \emph{not} improve performance; instead, it increases the per-frame time (FPS reduction) on both headsets for both videos (Fig.~\ref{img:cs2_stage_ms}). \frameworkname\ enables us to diagnose this counterintuitive result under standardized replay: since the visual stimulus and sensor inputs are held constant across runs, \frameworkname’s module-level logs let us localize where the slowdown occurs rather than attributing experimental confounds. The module-level timing breakdown (Fig.~\ref{img:cs2_stage_ms}) shows that the face detector dominates the per-frame budget in all conditions and that the low-precision stack increases time in the heaviest stages, yielding a higher end-to-end per-frame cost. This behavior is consistent with practical limitations of quantized execution in general-purpose inference backends (including OpenCV DNN): when optimized INT8 kernels and fusion are unavailable for the target CPU/operator set, quantized inference can incur additional conversion and data-movement overhead and may negate expected speedups~\cite{Bradski,onnxruntime}. Since our explicit PET uses OpenCV’s DNN (via OpenCV-for-Unity) on-device, we attribute the observed performance regressions to backend/kernel support.

% Candidate load affects the explicit PET in a way that complements Case Study~1~(\S\ref{casestudy1-exp1b}), where increasing candidate load primarily manifests as a performance degradation because the implicit PET’s per-frame work scales with the number of candidates processed. In this explicit PET, moving from the one-bystander to the two-bystander scripted video does not dramatically separate FPS within a headset (Fig.~\ref{img:cs2_fps}); instead, the more salient impact is on intent-to-enforcement behavior. Fig.~\ref{img:cs2_toggle_cond} shows that correctness becomes less stable under higher load and lower-precision settings, including a configuration where intended toggles fail entirely in the two-bystander video. This highlights an explicit-PET-specific stressor that standardized replay makes easy to test: when multiple candidates are present, the pipeline must not only detect gestures but also preserve the correct hand-face association over time, and failures here can prevent the PET from reaching the intended obfuscation state even while the pipeline continues to run.
Additionally, we find that the candidate load affects this explicit PET differently than the implicit BystandAR~(\S\ref{casestudy1-exp1b}). While moving from one to two bystanders does not strongly separate FPS within a headset, it instead stresses intent-to-enforcement behavior. Fig.~\ref{img:cs2_toggle_cond} shows that correctness becomes less stable under higher load and under the low-precision stack, including a configuration where intended protection toggles fail in the two-bystander video. 
% This highlights an explicit-PET-specific failure mode that \frameworkname\ makes easy to test: with multiple candidates present, the pipeline must maintain correct hand-face association over time for intent events to be enforced.

% Responsiveness follows the same pattern. Fig.~\ref{img:cs2_latency_proxy} shows that the time it takes for the pipeline to process frames with valid gestures increases under the low-precision stack and is consistently higher on MQ3 than ML2, aligning with the per-frame cost trends in Fig.~\ref{img:cs2_fps} and~\ref{img:cs2_stage_ms}. Under the standardized replay of identical intent events, \frameworkname\ therefore captures not only cross-headset throughput differences, but also how configuration choices translate into higher per-gesture processing cost at the moments when the PET must respond.

These results demonstrate \textbf{G2}: using \frameworkname's same end-to-end workflow as Case Study~1, \frameworkname\ supports a standardized evaluation of PETs across implicit and explicit designs by producing repeatable cross-headset performance measurements and design-appropriate PET outputs. 
% The combination of cross-headset ordering that matches Case Study~1, stage-attributed performance diagnosis, and controlled measurement of intent-to-enforcement behavior illustrates why \frameworkname’s workflow generalizes across PET categories while remaining faithful to what matters for each design.

\subsection{Case Study~3}\label{casestudy-iteration}

Case Studies~1 and~2 established that \frameworkname’s standardized workflow can produce repeatable measurements across headsets for both implicit and explicit PET designs under identical replayed inputs. In Case Study~3, we advance \textbf{G3} by demonstrating how \frameworkname\ enables PET design validation and iterative debugging by comparing proposed modifications exhibited by a PET under standardized replay of privacy-relevant edge cases. Such cases
present a key barrier in PET development: many failures occur only under brief occlusions, rapid motion, or tight spatial interactions between candidates. These conditions are difficult to recreate reliably in live trials and, therefore, difficult to debug or iterate on systematically.

\begin{figure*}[t!]
  \centering
  \captionsetup[subfigure]{justification=centering}

  \begin{subfigure}[t]{0.18\textwidth}
    \centering
    \includegraphics[width=\linewidth]{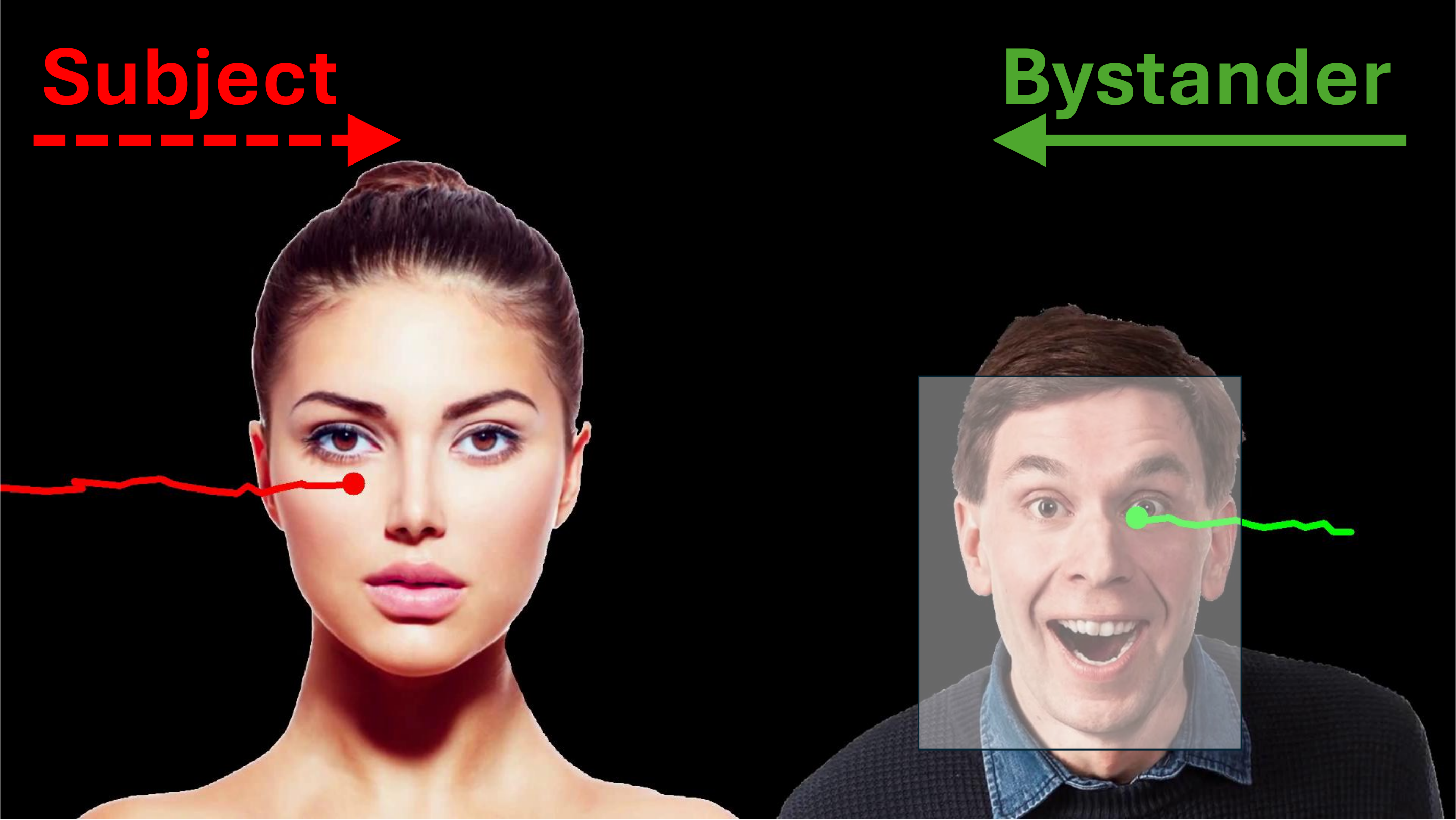}
    \caption{Before overlap\\correct tracking}
    \label{fig:exp3-before}
  \end{subfigure}\hfill
  \begin{subfigure}[t]{0.18\textwidth}
    \centering
    \includegraphics[width=\linewidth]{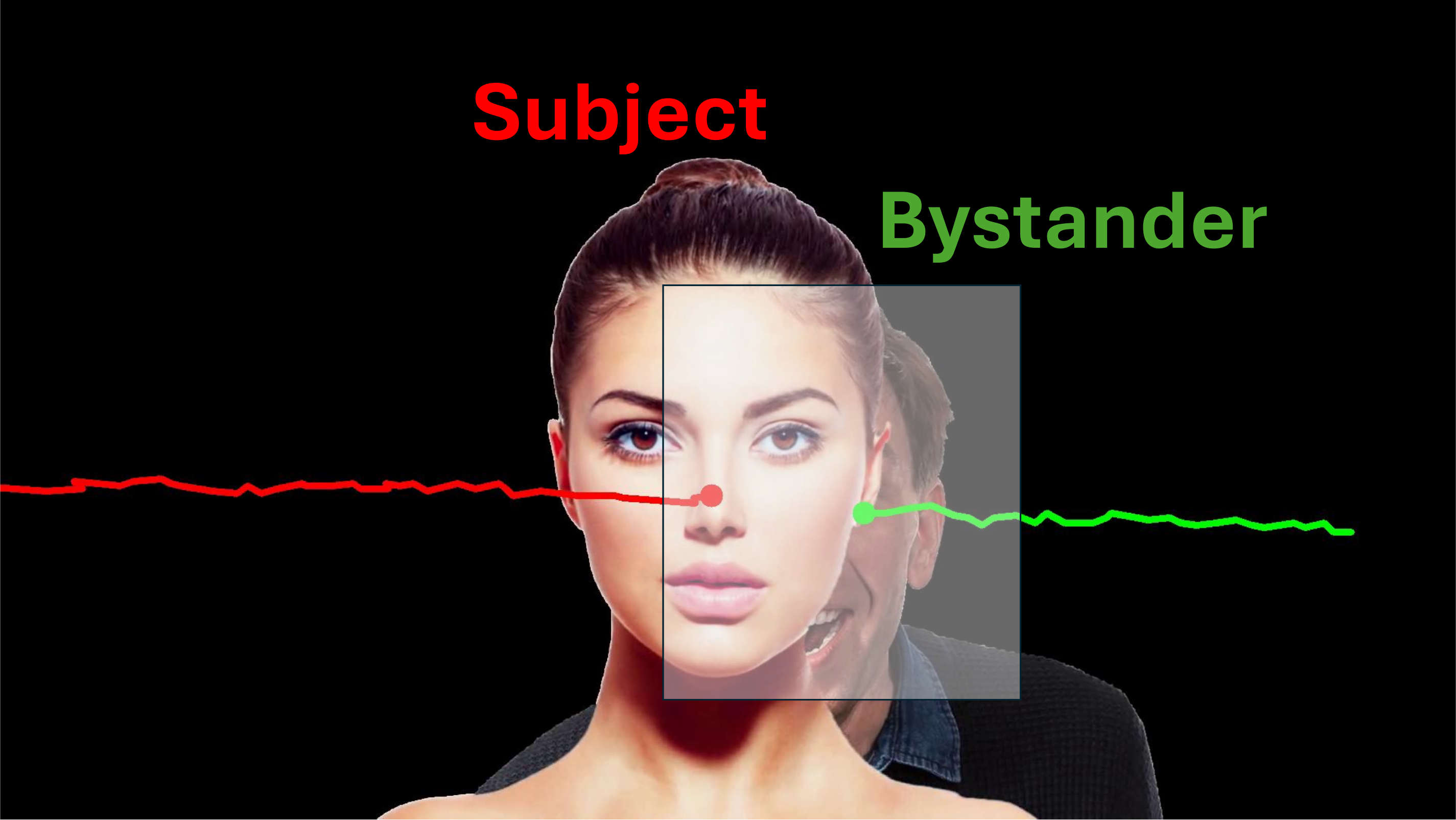}
    \caption{During overlap\\occluded face lost}
    \label{fig:exp3-during}
  \end{subfigure}\hfill
  \begin{subfigure}[t]{0.18\textwidth}
    \centering
    \includegraphics[width=\linewidth]{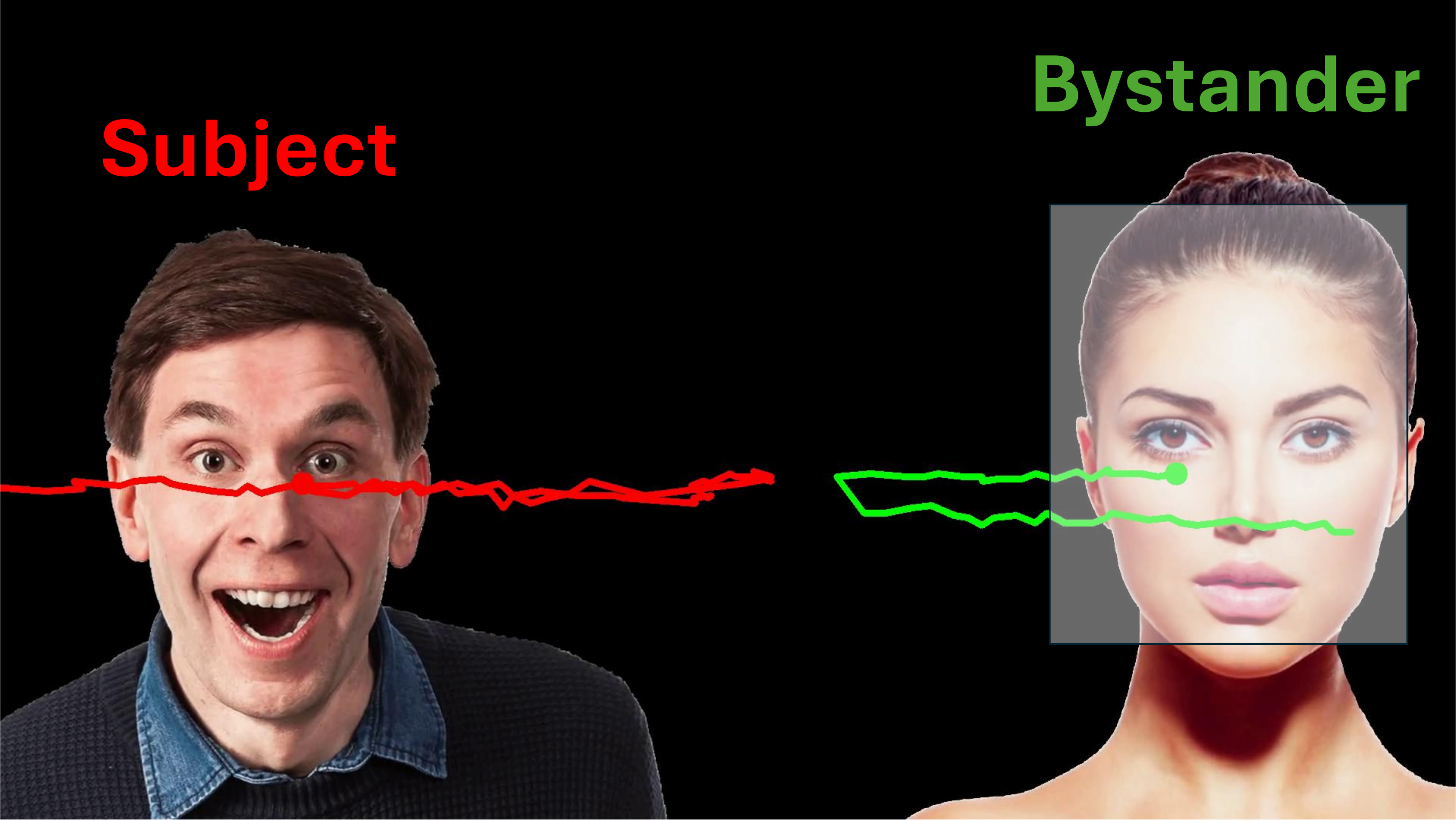}
    \caption{After overlap (F\textsubscript{s})\\detections swapped}
    \label{fig:exp3-after-fswap}
  \end{subfigure}\hfill
  \begin{subfigure}[t]{0.18\textwidth}
    \centering
    \includegraphics[width=\linewidth]{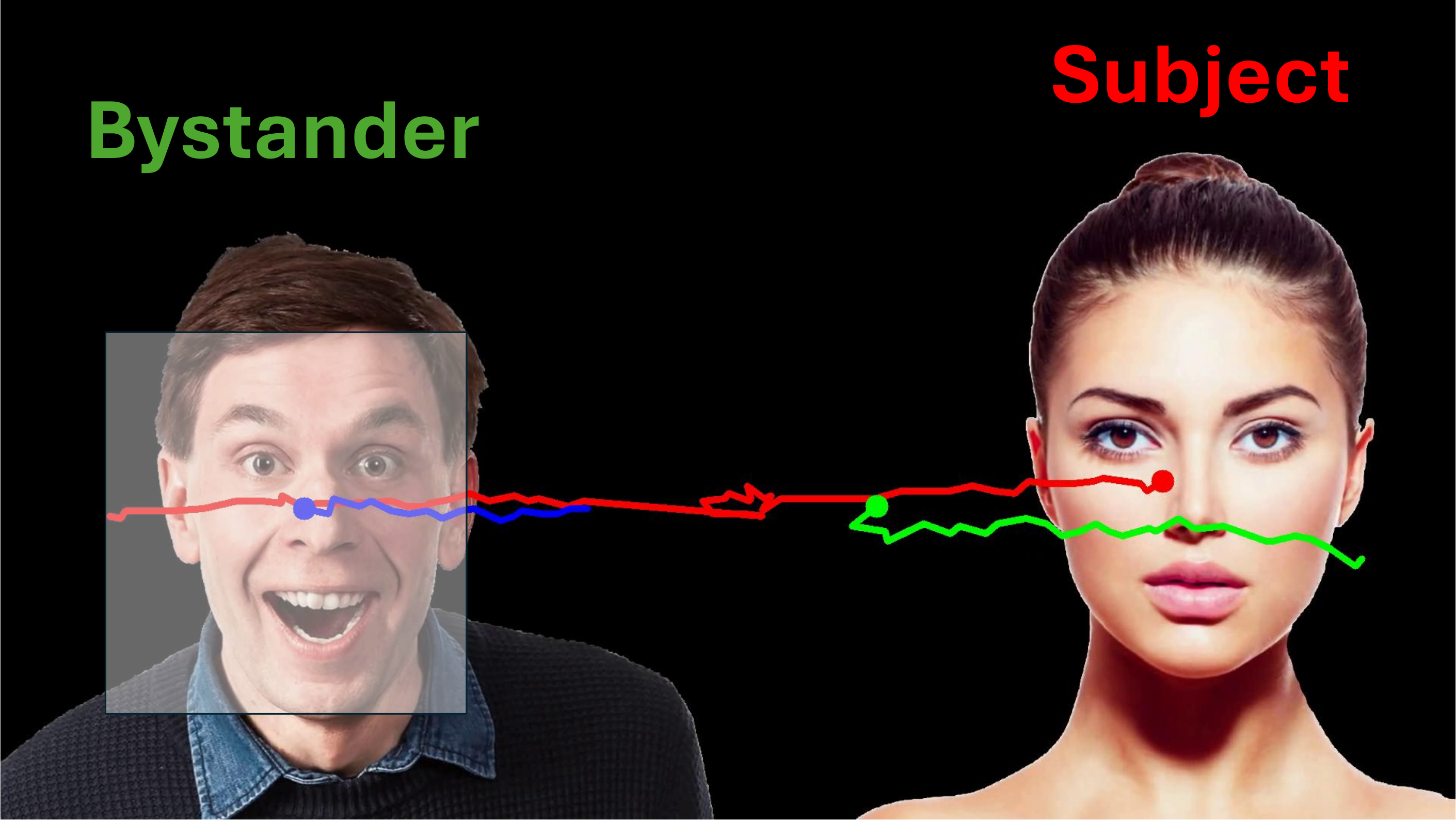}
    \caption{After overlap (F\textsubscript{l})\\occluded face re-detected}
    \label{fig:exp3-after-flost}
  \end{subfigure}\hfill
  \begin{subfigure}[t]{0.18\textwidth}
    \centering
    \includegraphics[width=\linewidth]{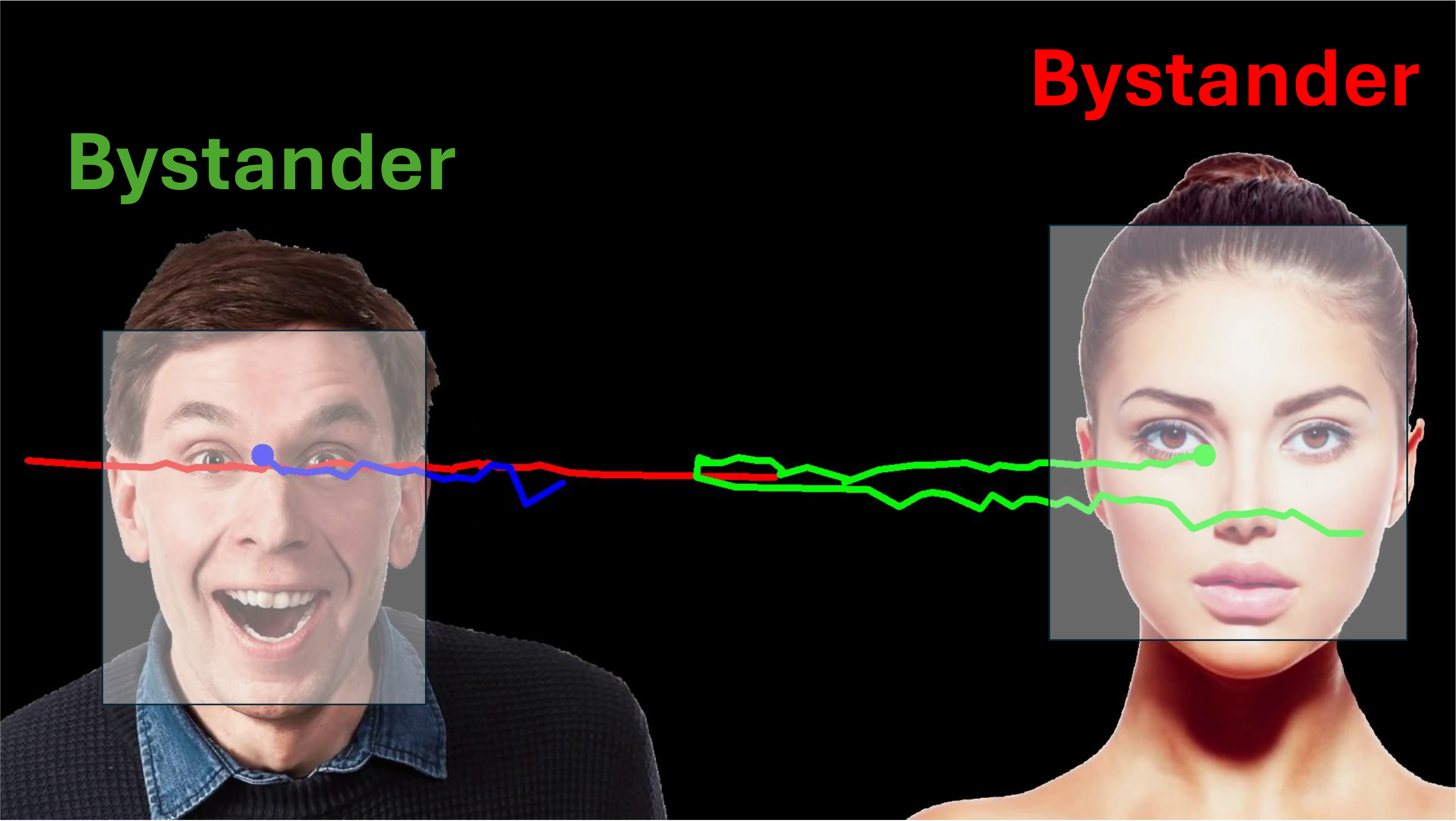}
    \caption{After overlap (F\textsubscript{d})\\one reassigned,\\one misassigned}
    \label{fig:exp3-after-fdrift}
  \end{subfigure}

  \caption{
  % Overlap outcomes for crossing faces: one ``Before'' and ``During'' panel followed by three representative ``After'' cases (F\textsubscript{s}, F\textsubscript{l}, F\textsubscript{d}). The bystander’s face is occluded by the translucent box.\
  Overlap outcomes for crossing faces: one ``Before'' and ``During'' panel, followed by three representative ``After'' cases for the observed failures. In F\textsubscript{s}, detections are swapped, causing loss of privacy and utility. In F\textsubscript{l}, the occluded face is re-detected as a new identity, breaking bystander protection. In F\textsubscript{d}, one face is reassigned while the other is misassigned, resulting in the loss of privacy, utility, or both, depending on the incorrect association. These cases show why overlap and occlusion are privacy-relevant edge cases for association logic. The bystander’s face is occluded by the translucent box.
  }
  \label{fig:exp3-failurescenarios}
\end{figure*}

\begin{table*}[b]
\caption{
% Case Study~3~(\S\ref{casestudy-iteration}) results of each BystandAR modification, and the baseline, for crossing and overlapping faces at varying speeds. Pass counts include all occurrences of P\textsubscript{s} and P\textsubscript{r} whereas Fail counts include all occurrences of F\textsubscript{s}, F\textsubscript{l}, and F\textsubscript{d}.
Case Study~3 results for the baseline and four BystandAR modifications across crossing and overlapping faces scenarios. KPP is the only modification that eliminates all observed privacy-relevant association failures across all scenarios.
}
  \label{tab:resultsexperiment3}
  \centering
  \resizebox{\textwidth}{!}{%
  \begin{tabular}{|l|cc|cc|cc|}
    \hline
    \multirow{2}{*}{BystandAR Variants} & \multicolumn{2}{c|}{Overlapping Faces} & \multicolumn{2}{c|}{Crossing Faces (Slow)} & \multicolumn{2}{c|}{Crossing Faces (Fast)} \\ \cline{2-3} \cline{4-5} \cline{6-7}
    & Pass & Fail & Pass & Fail & Pass & Fail \\
    \hline
    Baseline (unmodified) & 4 (3 P\textsubscript{s}, 1 P\textsubscript{r}) & 6 (6 F\textsubscript{s}) & 6 (1 P\textsubscript{s}, 5 P\textsubscript{r}) & 4 (4 F\textsubscript{s}) & 5 (5 P\textsubscript{r}) & 5 (4 F\textsubscript{s}, 1 F\textsubscript{l}) \\
    Naive Predicted Position (NPP) & 2 (1 P\textsubscript{s}, 1 P\textsubscript{r}) & 8 (8 F\textsubscript{s}) & 3 (3 P\textsubscript{s}) & 7 (7 F\textsubscript{s}) & 6 (5 P\textsubscript{s}, 1 P\textsubscript{r}) & 4 (3 F\textsubscript{s}, 1 F\textsubscript{d}) \\
    Closest Depth (CD) & 7 (6 P\textsubscript{s}, 1 P\textsubscript{r}) & 3 (2 F\textsubscript{s}, 1 F\textsubscript{l}) & 1 (1 P\textsubscript{s}) & 9 (9 F\textsubscript{s}) & 2 (2 P\textsubscript{s}) & 8 (6 F\textsubscript{s}, 2 F\textsubscript{d}) \\
    \textbf{Kalman Predicted Position (KPP)} & \textbf{10} (9 P\textsubscript{s}, 1 P\textsubscript{r}) & \textbf{0} & \textbf{10} (9 P\textsubscript{s}, 1 P\textsubscript{r}) & \textbf{0} & \textbf{10} (9 P\textsubscript{s}, 1 P\textsubscript{r}) & \textbf{0} \\
    Hybrid (KPP + CD) & \textbf{10} (10 P\textsubscript{s}) & \textbf{0} & 2 (2 P\textsubscript{s}) & 8 (8 F\textsubscript{s}) & \textbf{10} (10 P\textsubscript{s}) & \textbf{0} \\
    \hline
  \end{tabular}%
  }
\end{table*}

We return to BystandAR for this case study because it provides a suitable baseline for demonstrating \textbf{G3}. From Case Study~1, BystandAR’s pipeline executes stably and near real-time on modern headsets (especially ML2), which makes it feasible to run many controlled replays and attribute differences in outcomes to the intended algorithmic change rather than to unstable execution. In contrast, Case Study~2 showed that our OpenCV DNN-based gesture pipeline operates at single-digit FPS and is unstable on HL2; under such performance constraints, iteration is dominated by end-to-end feasibility and backend/runtime bottlenecks rather than by logic-level robustness. Therefore, we focus on BystandAR in this case study, which represents a mature implicit SOTA PET where standardized replay can be used to isolate and improve privacy-relevant failures.

\subsubsection{Experiment 3: Face association failures and proposed modifications}\label{casestudy3-exp3}

A privacy-preserving bystander PET must maintain a consistent mapping between each physical person and the PET’s internal identity/state over time; otherwise, the PET can obfuscate the wrong person (utility loss) or expose the person who should be protected (privacy loss). During preliminary runs of BystandAR under dynamic multi-person scenes~(\S\ref{casestudy1-exp1b}), we observed these failure cases when faces overlap or cross under occlusion within the visual stimuli. These scenarios stress BystandAR’s face association logic, which links new detections to previously tracked candidates. The baseline BystandAR's association logic uses Unity's \texttt{Physics.OverlapBox} to find overlapping facial detections across frames and select the first overlapping match to represent the same face across frames. This logic misassociates identities when faces overlap or when a face disappears briefly during occlusion.

We design this experiment around two IVs that together capture the iteration problem in \textbf{G3}. The first is the edge-case visual stimulus with three levels: $\textit{failure} \in \{\text{overlapping faces, } \allowbreak \text{slow crossing} \allowbreak \text{faces, } \text{fast crossing faces}\}$. The second IV is the association logic with five levels: $\textit{association} \in \{$baseline, Naive Predicted Position (NPP), Closest Depth (CD), Kalman Predicted Position (KPP), Hybrid (KPP+CD)$\}$. Apart from the baseline association logic, the other four logics represent our proposed modifications for the identified privacy-relevant association failures. NPP and KPP predict each detected face’s next location and match new detections to predicted positions using 3D Cartesian distance; KPP uses a Kalman filter to improve temporal stability and reduce sensitivity to jitter over NPP's naive approach of predicting position using the current and previous known locations. CD resolves overlaps by using depth estimates from BystandAR’s 2D-3D mapping to prefer the closest consistent candidate when bounding volumes overlap. The hybrid (KPP+CD) combines KPP and CD via a weighted score to handle both motion continuity and depth ordering. The algorithms for each modification can be found in Appendix~\ref{appendix-bystandar}.

The DV is the robustness of identity/state continuity under the replayed stimulus, reported as per-trial pass/fail and summarized across repetitions. We count a trial as a pass if the PET preserves a correct identity-to-privacy mapping across the scenario (including cases where tracking recovers without identity swap), and as a fail otherwise. We further use \frameworkname’s visual overlays to categorize failures into the three observable modes shown in Fig.~\ref{fig:exp3-failurescenarios}: swapped identities (F\textsubscript{s}), lost-and-recreated identity (F\textsubscript{l}), and drift/misassignment after occlusion (F\textsubscript{d}).

\textbf{Procedure.} We use three standardized stimulus videos, one for each level of the edge case IV, constructed from commercially licensed stock footage to produce controlled overlap/crossing occlusions. We run all conditions on ML2 to minimize hardware confounds and because Case Study~1 showed ML2 exhibits the most stable high-performance execution for BystandAR, making it suitable for repeated iteration runs. For each video, we replay the same recorded inputs and visual stimulus while swapping only the association logic inside the PET, repeating each logic ten times per video, resulting in a total of 150 trials. 
% Across repetitions, the replayed stimulus frames and the framework’s replay timing are held constant, so differences in outcomes reflect differences in association logic rather than differences in how the edge case is enacted.

\textbf{Results.} Table~\ref{tab:resultsexperiment3} summarizes outcomes across all association logics and edge cases. The baseline association fails frequently under overlapping and crossing faces, with failures dominated by identity swaps under overlap and by loss/reassignment under occlusion in crossing scenes. CD improves overlap handling (more passes under overlap than the baseline), but performs poorly under crossing faces where depth ordering is insufficient to preserve identity through occlusion. NPP does not consistently improve robustness, suggesting that naive motion continuation is too unstable under these occlusions. In contrast, KPP achieves perfect robustness across all three scenarios (10/10 passes in overlap, slow crossing, and fast crossing), indicating that stabilizing predicted motion with a Kalman filter is sufficient to resolve both overlap and occlusion cases. The hybrid strategy matches KPP under overlap and fast crossing but degrades under slow crossing, illustrating why controlled replay is valuable: two strategies that appear similarly effective in one edge case can diverge sharply under a slightly different motion regime.

Beyond aggregate pass rates, \frameworkname’s synchronized overlays make the failure mechanisms directly inspectable and comparable across proposed modifications (Fig.~\ref{fig:exp3-failurescenarios}). This is the practical iteration benefit enabled by standardized replay: once an edge case is captured in a stimulus, we can repeatedly reproduce it, localize the failure mode, and evaluate proposed modifications under identical conditions without re-running live trials or introducing variability from repeated enactments.

Taken together, these results demonstrate \textbf{G3}. \frameworkname\ makes privacy-relevant edge cases replayable and therefore testable as controlled inputs, enabling PET design validation and rapid iteration by swapping a single PET component (association logic) and comparing fixes under identical replayed stimuli. As the stimulus and replayed inputs are held constant across repetitions and algorithm variants, improvements in robustness can be attributed to the candidate fix itself rather than to uncontrolled differences in experimental execution.

%% file: TextFiles/New_Discussion.tex
% \vspace{-4mm}
\section{Discussion}\label{discussion}

In this section, we discuss \frameworkname\ capabilities demonstrated by our case studies~(\S\ref{casestudies}) and what their results imply for evaluating bystander PETs across headsets and PET designs. We also identify limitations that inform future work.

% \vspace{-3mm}
\subsection{Case Study~1}\label{discussion-cs1}
This case study demonstrates that \frameworkname\ can evaluate the same bystander PET under the same replayed inputs across different headsets. This capability is important because headsets differ in sensing pipelines and compute resources; without controlled replay, cross-device comparisons can be confounded by variation in how experimental trials are executed.

Beyond showing feasibility, the results highlight what cross-device reporting should capture when evaluating bystander PETs as real-time systems. In our setup, we vary candidate load and quantify how performance changes under standardized inputs. This enables a consistent characterization of whether a PET stays near real-time on each headset and how sensitive it is to increasing candidate load, using a shared logging format and analysis pipeline.

This case study also clarifies the practical value of \frameworkname’s workflow, fulfilling \textbf{G1}. The Data Collection and Data Replay stages provide a repeatable procedure with experimenter alignment, while the Visualization and Analysis stage produces synchronized experiment data and its overlays that make differences across headsets inspectable rather than anecdotal. To reduce the risk that \frameworkname\ inflates measured performance, our replay process disables any additional processing costs (visual marker detection once alignment is achieved), so the reported measurements better reflect the PET pipeline without being confounded by the framework.

% \vspace{-3mm}
\subsection{Case Study~2}\label{discussion-cs2}

This case study demonstrates \frameworkname’s ability to evaluate an explicit, gesture-driven bystander PET via the same workflow used to evaluate an implicit PET~(\S\ref{casestudy-performance}). Unlike implicit PETs, where protection decisions are derived continuously from sensed context, explicit PETs introduce discrete intent events (e.g., opt-in/opt-out gestures) that command state changes. Therefore, evaluation must capture not only system performance, but also whether intent events are (i) recognized, (ii) associated with the intended face candidate, and (iii) reflected as the correct protection state transition over time, including the intent-to-enforcement delay.

Hence, evaluation of explicit PETs additionally requires intent-to-state correctness and intent-to-enforcement responsiveness. 
% This case study shows that \frameworkname\ supports these measurements within the same end-to-end pipeline: replay provides standardized stimuli and repeatable inputs so the explicit PET can be executed consistently across runs, and analysis uses the logged outputs (recognized gestures, associations between gesture and face candidates, and per-candidate protection state over time) to compute correctness and responsiveness, making it possible to quantify both whether intent is applied to the intended candidate and how quickly the system reacts.
This case study shows that \frameworkname\ supports these measurements within a single end-to-end pipeline. During replay, \frameworkname\ provides standardized stimuli and repeatable inputs, allowing the explicit PET to run consistently across trials. During analysis, we use logged outputs (recognized gestures, gesture–face associations, and each candidate’s protection state over time) to compute correctness and responsiveness. Together, these measures quantify whether intent is applied to the intended candidate and how quickly the system reacts.

A key result from this case study is that the low-precision (INT8) model stack does not improve performance; instead, it reduces FPS on each headset in our experiment~(\S\ref{casestudy2-exp2}). This outcome is counterintuitive if one assumes quantization reliably reduces inference cost. \frameworkname\ helps explain this effect because it logs module-level timings in addition to end-to-end FPS. Under standardized replay, the runtime breakdown shows that the face detector dominates the per-frame budget, and the low-precision stack increases (rather than decreases) time in the heaviest stages, yielding a higher end-to-end per-frame PET cost. This behavior is consistent with practical limitations of quantized execution in general-purpose inference backends (including OpenCV’s DNN module): if the backend cannot fuse quantization/dequantization patterns or lacks optimized INT8 kernels for the target CPU and operator set, quantized inference can incur extra conversion and data-movement overhead and may even fall back to higher-precision implementations, eliminating the expected speedup. Hence, the high-precision model stack in our experiment for this case study results in better performance.

Taken together with Case Study~1, this case study completes \textbf{G2}: 
% \frameworkname is not tied to a single PET category or a single set of metrics; instead, it standardizes the \emph{workflow} and the input/output logging needed for replay, while permitting PET-appropriate dependent variables for each design.
\frameworkname\ standardizes the evaluation workflow across PET designs, while allowing the dependent variables to differ based on whether the PET is implicit or explicit.

% \vspace{-3mm}
\subsection{Case Study~3}\label{discussion-cs3}

This case study fulfills \textbf{G3} by demonstrating how \frameworkname\ supports PET design validation and iterative debugging by capturing challenging situations once and replaying them to compare PET variants under identical inputs. Many privacy-relevant failures occur briefly (e.g., when candidates overlap or move quickly) and are difficult to reproduce consistently with ad-hoc live experimental trials. By converting such a failure into a replayable trial, \frameworkname\ enables more grounded diagnosis of privacy-relevant failures and more reliable comparison of their fixes.

In this case study, Data Replay allows us to test alternative association logic on the same captured inputs and to attribute changes in outcomes to the PET modification rather than to differences in trial execution. The synchronized overlays further help interpret why a failure occurred by making state continuity and identity association issues visible during replay, connecting quantitative outcomes to observable failure modes.

This case study shows that \frameworkname\ is not only an evaluation harness, but also a practical mechanism for improving SOTA PETs. Using \frameworkname, we were able to implement and validate a stronger PET variant of the current SOTA implicit PET (BystandAR) by demonstrating clearer state continuity under the same captured failure case, supported by the framework’s synchronized logs and overlays. This ability to move from observing a failure to shipping a verified improvement is a central advantage of \frameworkname, and it is difficult to achieve with prior ad-hoc evaluation approaches that lack reproducability.

% \vspace{-3mm}
\subsection{Ethical Considerations}\label{discussion-ethical}
\frameworkname\ has several ethical implications for PET research. It supports studying privacy-sensitive scenarios in an ethically responsible manner via video recording-based replay instead of repeatedly exposing participants to such environments in early-stage evaluations. Moreover, as the framework supports PET design validation, debugging, and privacy-performance profiling under controlled conditions, researchers can identify and fix flaws in their PETs before moving to later-stage human-subject studies. Hence, \frameworkname\ protects participants' privacy by limiting their exposure to immature PET designs. In addition, \frameworkname's responsible usage depends on working with consented, licensed, or synthetic stimuli, and on careful collection, storage, and sharing of replayable visual data that may contain identifiable bystanders or sensitive contextual information. Hence, we do not redistribute raw scenario videos. Instead, we provide the framework code, analysis scripts, and share the visual stimuli sources for obtaining or recreating the stimuli under the appropriate consent and licensing terms.
% \bdj{could this be stronger? In that we don't share the dataset of videos at all, but provide links such that other researchers can access them with the appropriate license? Not sharing stock footage is more likely a copyright violation, and I don't think that is the entire ethical risk you're referring too.}\bo{agreed}\ibrahim{revised; please see if the comment is addressed now.}

% \vspace{-3mm}
\subsection{Limitations and Future Work}\label{discussion-limitations}

Our evaluation has several limitations that motivate future work.

\frameworkname’s pipeline is not restricted to visual stimuli and can log and replay additional modalities (including audio). However, our evaluation of the framework focuses only on vision-based bystander PETs. Future work can broaden the scope of this evaluation by instantiating and benchmarking audio-driven (or multimodal) PETs within the same record-replay workflow using realistic acoustic conditions (e.g., crowd noise) alongside the visual stimulus.

% \frameworkname’s record-replay workflow prioritizes reproducibility: the same inputs can be replayed across headsets and configurations to isolate PET behavior and performance under controlled conditions. The trade-off is that replayed stimuli do not capture social dynamics present with live bystanders (e.g., mutual adaptation, conversational flow, or behavior changes in response to privacy cues). \frameworkname\ is therefore a complement to—not a replacement for—human-subject evaluations. Studies centered on user experience or on developing new PET interactions should pair \frameworkname\ with in-situ recordings and/or live deployments while retaining the same logging and replay interfaces.
\frameworkname’s record-replay workflow prioritizes controlled replication of evaluation conditions to meet the goals of early-stage PET evaluation.
% , such that the same inputs can be replayed across headsets and trials to isolate PET behavior, performance, and privacy-performance trade-offs. 
The trade-off of this controlled evaluation is that replayed stimuli do not capture social dynamics present with live bystanders (e.g., conversational flow, or behavior changes in response to privacy cues). \frameworkname\ therefore complements later-stage human-subject evaluations focusing on ecological validity, user experience, and social acceptability.

% Additionally, this paper primarily focused on performance and privacy-relevant failure modes, with limited standalone privacy-quality and usability evaluation. Although the framework supports hooks for ground-truth-based evaluation of privacy protection and visual outputs for usability studies, full integration and validation of these dimensions remain future work.
This work focuses on performance and privacy-performance trade-offs in early-stage PET evaluation, instead of standalone privacy-protection or usability evaluation. However, \frameworkname's design can support both in future work. The framework’s outputs can log arbitrary PET output data in a standardized format, including obfuscation states and bystander labels for each frame. In our case studies, we already log privacy-relevant PET outputs, including per-frame obfuscation states and bystander labels for each detection, to support the privacy-related analyses in Case Studies~2 and~3. Future studies can extend these logged outputs to conduct in-depth privacy-protection analysis using more comprehensive or task-specific protection metrics. Similarly, the synchronized overlays of PET outputs on the visual stimuli generated by the framework can support usability evaluation of PET behavior. 
For example, large-scale crowd-sourced surveys of participants’ perceptions of the PET outputs could be enabled using the framework's visualizations, similar to I-Pic's usability evaluation via an online survey~\cite{aditya2016pic}.

Lastly, our experiments did not include AR content rendering. In real-world conditions, headsets render AR content, such as ubiquitous glanceable interfaces~\cite{lu2023wild,lu2024did},  alongside the PET, which may affect performance. Our framework is capable of analyzing performance alongside standard rendering workloads. Future studies can validate PET behavior under rendering workloads (e.g., peripheral interfaces and gaming).

%% file: TextFiles/New_Conclusion.tex
% \vspace{-3mm}
\section{Conclusion}\label{conclusion}

Reproducible cross-device evaluation of behavior, performance, and privacy-performance trade-offs of visual bystander PETs for AR headsets is challenging due to limitations of the existing PET evaluation pipeline. We presented \frameworkname, a framework that enables low-overhead early-stage evaluation of bystander PETs across devices by standardizing (1) PET input/output, (2) visual stimuli, and (3) the record-replay of PET-consumed sensor data.
Across three case studies on HL2, ML2, and MQ3, we demonstrated that \frameworkname\ produces comparable cross-device performance measurements for a PET under controlled conditions, supports both implicit and explicit PET designs, and enables rapid prototyping by replaying privacy-relevant edge cases to validate proposed modifications, demonstrating an improved variant over a SOTA PET. 
These results show how \frameworkname\ supports low-overhead, repeatable design validation, debugging, and privacy-performance profiling across devices and PET designs, helping move PET development beyond ad-hoc, device-specific evaluation toward a more reproducible workflow for emerging AR systems.